\documentclass{article}

\usepackage{authblk}
\usepackage[utf8]{inputenc}
\usepackage[left=2cm,right=2cm,top=1.5cm,bottom=1.5cm]{geometry}
\usepackage{amsmath}
\usepackage{graphicx}
\usepackage{subfig}
\usepackage{mathtools}
\usepackage{mathpazo}
\usepackage[mathpazo]{flexisym}
\usepackage{breqn}
\usepackage{xcolor}
\usepackage{comment}

\graphicspath{{figs/}}

\usepackage[style=numeric-comp,style=nature,url=false]{biblatex}
\DeclareMathOperator{\sech}{sech}

\title{Kink-inhomogeneity interaction in the sine-Gordon model }
\author[1]{Jacek Gatlik}
\author[1]{Tomasz Dobrowolski}
\author[2]{Panayotis G. Kevrekidis}
\affil[1]{\textit{Pedagogical University of Krakow, Podchor\c{a}\.zych  2, 30-084 Cracow, Poland}}
\affil[2]{\textit{Department of Mathematics and Statistics, University of Massachusetts, Amherst, Massachusetts 01003-4515, USA}}
\date{\today}

\begin{document}
\maketitle

\begin{abstract}
In the present study the interaction of a sine-Gordon kink with a localized inhomogeneity is considered. In the absence of dissipation, the inhomogeneity considered is found to impose a potential energy barrier. The motion of the kink for near-critical values of velocities separating transmission from barrier reflection is studied. Moreover, the existence and stability properties of the kink at the relevant saddle point are examined and its dynamics is found to be accurately captured by effective low-dimensional models. In the case where there is dissipation in the system, below the threshold value of the current, a stable kink is found to exist in the immediate vicinity of the barrier. The effective particle motion of the kink is investigated obtaining very good agreement with the result of the original field model. Both one and two degree-of-freedom settings are examined with the latter being more efficient than the former in capturing the details of the kink motion.
\end{abstract} \hspace{10pt}

\section{Introduction}
The sine-Gordon (sG) model originally appeared in the description of surfaces of constant negative curvature embedded in three-dimensional space. This equation constitutes the Gauss-Codazzi integrability condition of the surface~\cite{mclachlan}. Primarily, the model was introduced to physics in the context of the studies on crystal dislocations \cite{Frenkel1939}. Since then, it has found many applications in describing a variety of physical systems~\cite{Kivshar1989,CKW14}. 

One of the prototypical examples showcasing the relevance of the sG model concerns its application to quasi-one-dimensional ferromagnetic materials with an easy plane anisotropy and their behaviour in an external magnetic field \cite{Zharnitsky1998}. Experimental studies of this system confirm the main theoretical predictions \cite{Kijems1978,Jongh1981,Steiner1983}. Also, the relevant system has been successfully leveraged to describe ferroelectrics~\cite{Dodd1984,Pouget1984,Pouget1985,Bugaychuk2002}. Moreover, the orientation angle of the molecules in liquid crystals has been argued to satisfy an overdamped and externally driven sine-Gordon equation~\cite{Lam1992}. An additional example where the sG model (especially in its damped-driven variant) has been shown to be experimentally accessible concerns an array of coupled torsion pendula; see for a relatively recent demonstration the experiments of~\cite{lars}.
 
Arguably, the most widespread application of the sG model concerns the description of a device called the Josephson junction that emerged as a result of the so-called Josephson effect~\cite{Josephson1962}. Predictions of this work found experimental confirmation a year later~\cite{Anderson1963}. Josephson junctions (JJs) have been thoroughly studied over the years \cite{Barone1982,Malomed2014} and have found numerous practical applications \cite{Braginski2019}. In order to obtain the most realistic description of the JJs, additional terms were introduced describing the dissipation due to tunneling of normal electrons across the barrier, the dissipation caused by the flow of normal electrons parallel to the barrier and moreover the bias current~\cite{Malomed2014}. Additionally, in the context of condensed matter physics, the presence of inhomogeneities in the form of ``impurities'' is a fairly common feature. More concretely, in the JJ setting, the typical inhomogeneities are microshorts which are local regions of high Josephson current \cite{Scott1978,Golubov1988,Kivshar1989}. The effect of modulation of the thickness of a dielectric layer separating the two superconducting electrodes has been described in many different ways \cite{Mkrtchyan1979,Malomed1990,Dobrowolski2020}. Another way in which explicitly position-dependent functions enter the sine-Gordon model is presented in the works~\cite{Demirkaya2013,Kevrekidis2014,Demirkaya2014}. The latter possibility has been motivated by the widespread relevance of ${\cal PT}$-symmetric systems in optical, as well as more generally in dispersive wave systems~\cite{jyang}.

A considerable volume of work has also focused on the effect of shape deformation of the junction on its properties \cite{Benabdallah1996,Kemp2002,Kemp2002b,Gulevich2006,Caputo2014,Monaco2016}. In this approach, some modifications of the junction shape are proposed in order to obtain its desired properties. In particular, the influence of the curvature on the dynamics of the gauge invariant phase difference between two superconducting electrodes that comprise the JJ was studied in \cite{Dobrowolski2012,Gatlik2021}. The equation that describes this system was obtained on the basis of field dynamics governed by Maxwell's equations in the insulator and London's equations in superconducting electrodes with Ginzburg-Landau current of Cooper pairs. The description in this case agrees with the same result obtained on a purely geometrical background as a consequence of the geometrical reduction of the sine-Gordon model to a lower dimensional curved subspace \cite{Dobrowolski2009}.
 
In the present work, we focus on describing the interaction of a kink-like effective particle in the sG model with inhomogeneities for initial velocities close to the critical velocity. This choice of initial conditions can render the interaction time significantly longer close to this critical point which highlights all aspects of the interaction. Our interest lies in systematically describing this interaction via a low-dimensional, effective-particle approach, both for the Hamiltonian (conservative) but also for the dissipative partial differential equation (PDE) setting. In addition to exploring the relevant PDE dynamics, emphasis is placed on effective, low-dimensional descriptions of the solitary wave in the corresponding energy landscape. In the next section (section 2) we will describe the field model to be studied. We determine the shape of the kink both in the absence and in the presence of dissipation and external forcing in the system. This section also examines the linear stability of the solutions. Section 3 is devoted to effective descriptions of different dimensionalities (one- and two-degree of freedom approaches) and the limits of their applicability, as well as the comparison between them, as well as with the original PDE. The last section contains our conclusions and a number of proposed directions for possible future study.

\section{System description}
In the present article, in line with the above discussion, we study the perturbed sine-Gordon model of the form:
\begin{equation}
\label{sine-gordon}
\partial_t^2 \phi + \alpha \partial_t \phi - \partial_x (\mathcal{F}(x)\partial_x \phi) + \sin \phi = -\Gamma ,
\end{equation}
where the function $\mathcal{F}(x)$ represents the inhomogeneity.
More specifically, our motivation for considering this type of modification stems from the need to take into account the curvature in the description of the long Josephson junction. The detailed physical considerations leading to this effective equation are presented in the earlier works of~\cite{Dobrowolski2012,Gatlik2021}. The same equation can be obtained from the mathematical procedure of projecting the sine-Gordon equation defined in a flat 3-dimensional space, into a 1-dimensional subspace, non-trivially embedded in the initial space~\cite{Dobrowolski2009}. In the above equation $\alpha$ is the dissipation coefficient while $\Gamma$ represents a constant external forcing. In the context of a Josephson junction, the constant $\Gamma$ is interpreted as a bias current. In the absence of dissipation and external forcing, the total energy is conserved. For later convenience, we separate the function $\mathcal{F}(x)$ into a part describing the unperturbed system and a term describing its disturbance $g$ 
\begin{equation}
\label{inhomogeneity}
\mathcal{F}(x) = 1 +\varepsilon g(x).
\end{equation}
The parameter $\varepsilon$ controls the magnitude of the perturbation. We will assume that this parameter is small.  

\subsection {The non-dissipative case with $\alpha=0$ and $\Gamma=0$}
First, we will  focus on describing the simplified case, i.e., one in which the constants $\alpha$ and $\Gamma$ are equal to zero. 
 Although the energy of a free kink in a homogeneous system (featuring distinct asymptotic equilibria) corresponds to a minimum of energy described by the formula
\begin{equation}
\label{energy_kink}
E =\int_{-\infty}^{+\infty}dx\left[\frac{1}{2}(\partial_t\phi)^2+\frac{1}{2}(\partial_x\phi)^2+(1-\cos\phi)\right],
\end{equation}
it can be further lowered in the presence of inhomogeneity. The total energy of the arbitrary field configuration in a heterogeneous system is of the form
\begin{equation}
\label{energy}
E_H =\int_{-\infty}^{+\infty}dx\left[\frac{1}{2}(\partial_t\phi)^2+\frac{1}{2}\mathcal{F}(x)(\partial_x\phi)^2+(1-\cos\phi)\right].
\end{equation}
We describe the process of interaction of the kink with admixture present in the system. In this section, the function $g$ is taken in the form $g(x) = \tanh(x)-\tanh(x-L)$. In this formula, $L$ defines the width of the inhomogeneity.

It is relevant here to briefly discuss the method with which these profiles are obtained. We have utilized a Newton-Raphson iteration which, through its quadratic convergence, has ensured the rapid identification of the relevant kink profiles. The steady state problem is discretized by means of centered finite-differences (of second order) and the accuracy of the findings has been ensured by means of discretizations of different spacing $\Delta x$. It should be added that as part of the Newton-Raphson procedure, we also construct the Jacobian evaluated at the kink profile. This, upon convergence, provides us with the linearization matrix of the relevant problem that will be used for the numerical identification of the eigenfrequencies $\omega$ discussed in more detail in what follows.

\begin{figure}[ht]
    \centering
    \subfloat{{\includegraphics[height=5cm]{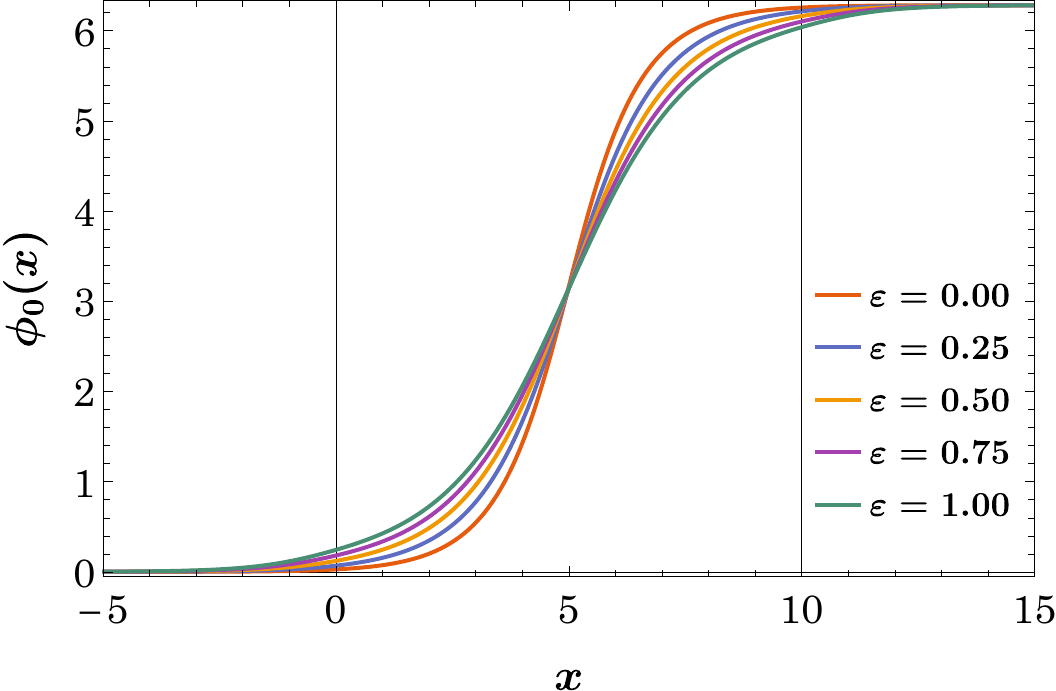}}}
    \quad
    \subfloat{{\includegraphics[height=5cm]{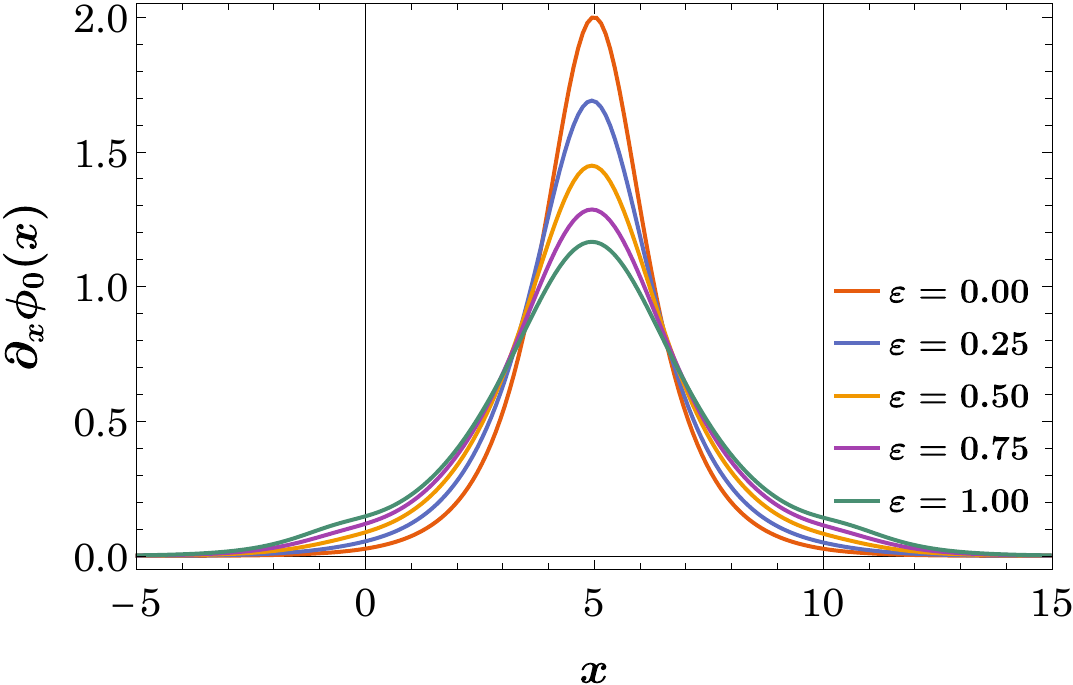}}}
    \caption{Profiles of static solutions for different values of $\varepsilon$ (left figure) and, for a better visualisation, gradients of static configurations (right figure).  In all cases, the size of the inhomogeneity is $L = 10$. The inhomogeneity in the figures is located between the vertical lines for $x=0$ and $x=L$.}
    \label{fig_01}
\end{figure}

\begin{figure}[ht]
    \centering
    \subfloat{{\includegraphics[height=5cm]{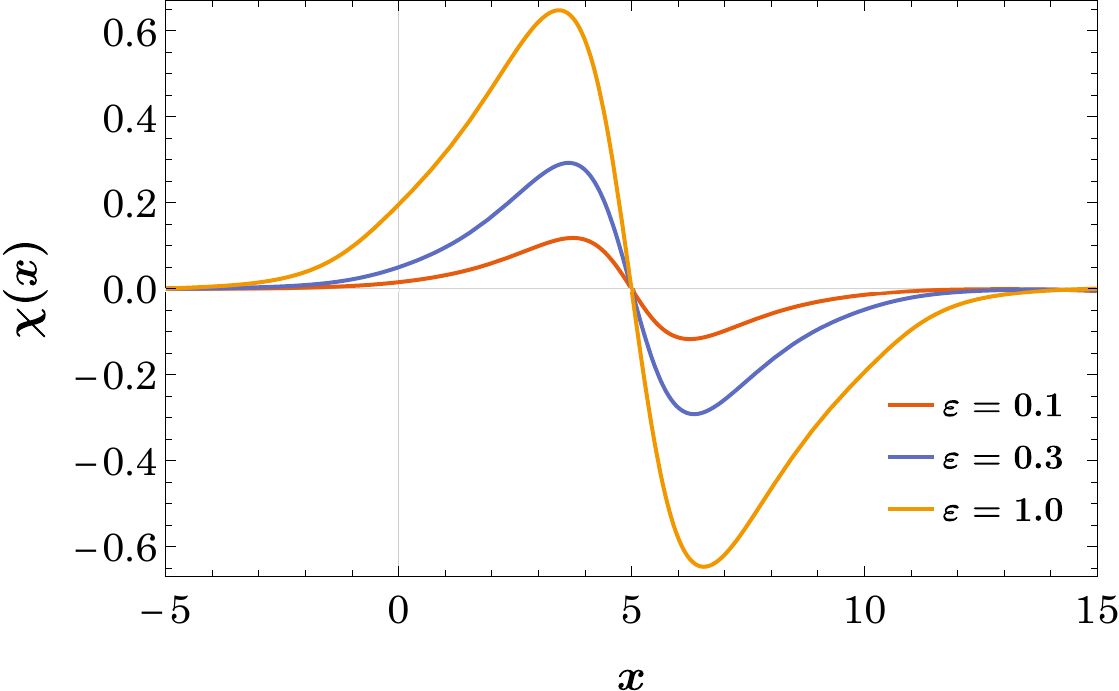}}}
    \quad
    \subfloat{{\includegraphics[height=5cm]{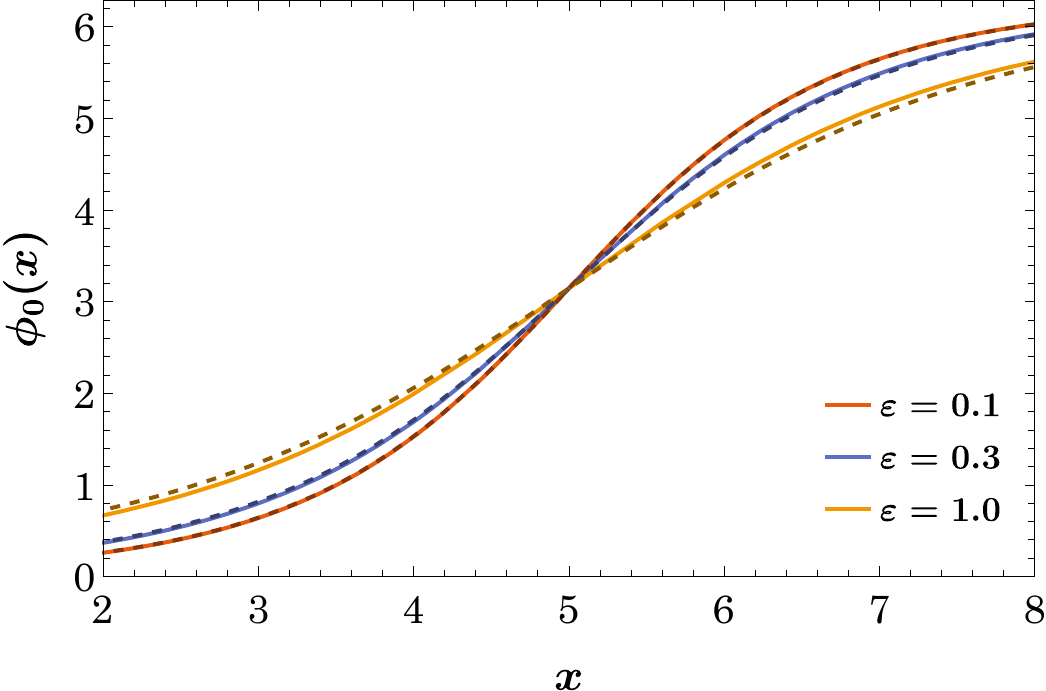}}}
    \caption{On the left panel, the $\chi(x)$ value determined from Eq. \eqref{linear-2} is shown 
    for different values of $\varepsilon$. On the right
    panel, the solid line shows the sum of the kink ansatz $\phi_K(x)$ and $\chi(x)$ value for different $\varepsilon$, while the dashed line is the corresponding static solution as determined from Netwon's method.}
    \label{fig_02}
\end{figure}

The deformation of the kink profile as a function of $\varepsilon$ is evident in Fig.~\ref{fig_01} showcasing the widening of the kink as $\varepsilon$ is increased. From a more quantitative perspective, the deformation of the kink profile for small $\varepsilon$ and the stability of this configuration can be examined in the framework of a linearized approximation. To begin with, we assume that the field $\phi$ is a slightly perturbed kink solution of the model (\eqref{sine-gordon} with $\alpha=0$ and $\Gamma=0$). We insert the decomposition $\phi(t,x) = \phi_0(x) + \psi(t,x)$ into equation \eqref{sine-gordon} obtaining, up to linear terms in the $\psi$ correction, the equation
\begin{equation}
\label{linear-1}
\partial_t^2 \psi - \partial_x \left( {\cal F}(x)\partial_x \psi\right) + ( \cos \phi_0 ) \psi = 0.
\end{equation}
At this point we would like to emphasise that $\phi_0(x)$ can be decomposed into static kink $\phi_K$ of the sine-Gordon model and   a time independent correction $\chi$ depending also on the geometry of the system i.e. $\phi_0(x) = \phi_K(x)+\chi(x)$. This is intended to capture the steady state solution of the perturbed (in the presence of the inhomogeneity) problem. In particular, for small values of the parameter $\varepsilon$, the correction $\chi(x)$ can be calculated from the equation:
\begin{equation}
\label{linear-2}
    -\partial_x \left( {\cal F}(x)\partial_x \chi \right) + (\cos \phi_K ) \chi = \varepsilon\partial_x\left( g(x)\partial \phi_K\right).
\end{equation}
This equation describes the time-independent deformation, which is uniquely determined by the function describing the inhomogeneity and the analytical form of the underlying solution. The solutions of the equation \eqref{linear-2} for different values of $\varepsilon$ are presented in the figure \ref{fig_02}. It is clear that the relevant contributions are antisymmetric along the (former) direction of the translational invariance of the homogeneous model kink and upon addition to the homogeneous static kink, they modify its effective width. Moreover, the profiles of the static (numerically exact up to a prescribed tolerance) solutions $\phi_0(x)$ obtained from the perturbed sine-Gordon model \eqref{sine-gordon} are compared with the function $\phi_K(x)+\chi(x)$, where $\phi_K(x)$ is static kink solution of the sine-Gordon model in the homogeneous case. The figure shows that, even for $\varepsilon=1$, there is little difference between the solution derived from equation \eqref{linear-2} and the numerical solution derived from the field model \eqref{sine-gordon}.

\begin{figure}[ht]
    \centering
    \subfloat{{\includegraphics[height=5cm]{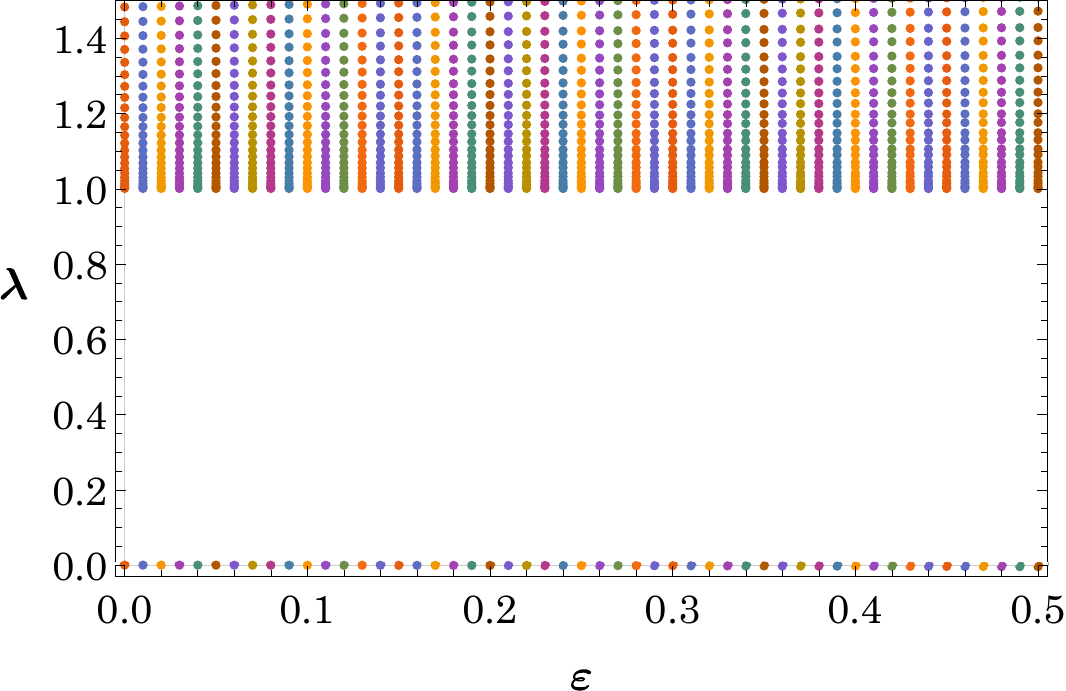}}}
    \quad
    \subfloat{{\includegraphics[height=5.1cm]{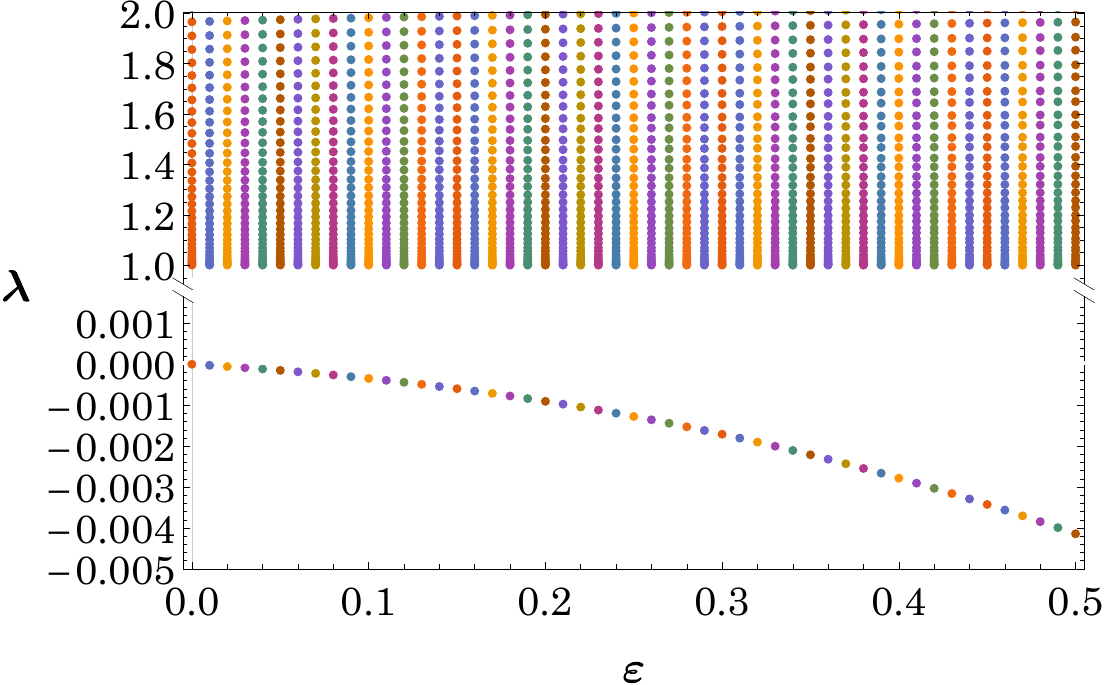}}}
    \caption{Squared eigenfrequencies $\lambda=\omega^2$ calculated for the quasi-static configuration depending on the value of $\varepsilon$ for $L=10$.
    See also the discussion in the text.}
    \label{fig_03}
\end{figure}

On the other hand, the equation \eqref{linear-1} contains information about the time dependent perturbation of the underlying solution. We then adopt a particular form of the time dependence of the function $\psi$ i.e. $ \psi(t,x) = e^{i \omega t} v(x) .$ This standard approach allows us to examine the spectral stability of the underlying configuration $\phi_0$
\begin{equation}
\label{linear-3}
 - \partial_x \left( {\cal F}(x)\, \partial_x v(x)\right) + (\cos \phi_0 ) \, v(x) = \lambda v(x),
\end{equation}
where $\lambda=\omega^2$. By abbreviating the left-hand side of the last equation $\hat{{\cal L}} \, v(x) $ we obtain the eigenequation for the linearization operator $\hat{{\cal L}}$
\begin{equation}
\label{linear-4} \hat{{\cal L}} \, v(x) = \lambda v(x) \, .
\end{equation}
The spectrum of the operator $\hat{{\cal L}}$ obtained from the equation \eqref{linear-4} consists of a continuous spectrum and a discrete negative value (see Figure \ref{fig_03}). The latter eigenvalue pertains to the previously vanishing eigenfrequency (of the homogeneous limit) associated with the translational invariance of the homogeneous problem. In the present case, the negative associated squared eigenfrequency corresponds to a real eigenvalue illustrating that the relevant static configuration corresponds to an unstable equilibrium, more specifically a saddle point of the (undamped, non-driven) Hamiltonian limit of the system. This, in turn, represents a potential energy maximum of the effective energy landscape, whose energy we expect to separate between the transmission dynamics (for energies higher than that of this configuration) and the reflection features (for energies below those of this maximum).


\subsection{The case with dissipation and external forcing $\alpha  \neq 0$ and $\Gamma \neq 0$}
For $\Gamma \neq 0 $, similarly as in the previous case, during interaction with the inhomogeneity the kink may or may not go over the barrier. The reflection in this case is much more interesting than for $\Gamma=0$. The physical cause is the presence of a constant force pressing the particle against the potential barrier. For this reason, after the bounce, as we will see in the dynamical simulations below, the kink is again pushed towards the barrier. The presence of dissipation makes subsequent reflections smaller and smaller, until finally the kink stops at a certain distance from the barrier. The form of this resulting static configuration is specified by the equation
\begin{equation}
\label{sine-gordon-static}
 - \partial_x (\mathcal{F}(x)\partial_x \phi_0) + \sin \phi_0 = -\Gamma .
\end{equation}
In this equation, the bias current must be less than the threshold value. The current threshold value separates the current values for which kink is stopped before the barrier from the values for which kink overcomes the barrier.

\begin{figure}[ht]
    \centering
    \subfloat{{\includegraphics[height=5cm]{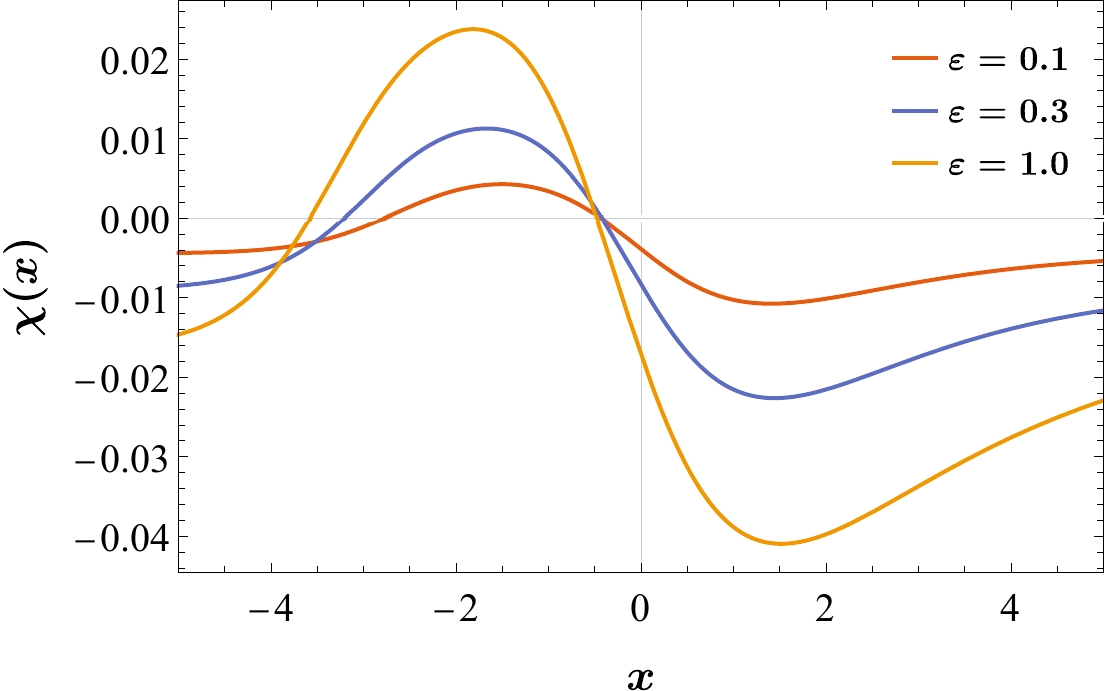}}}
    \quad
    \subfloat{{\includegraphics[height=5cm]{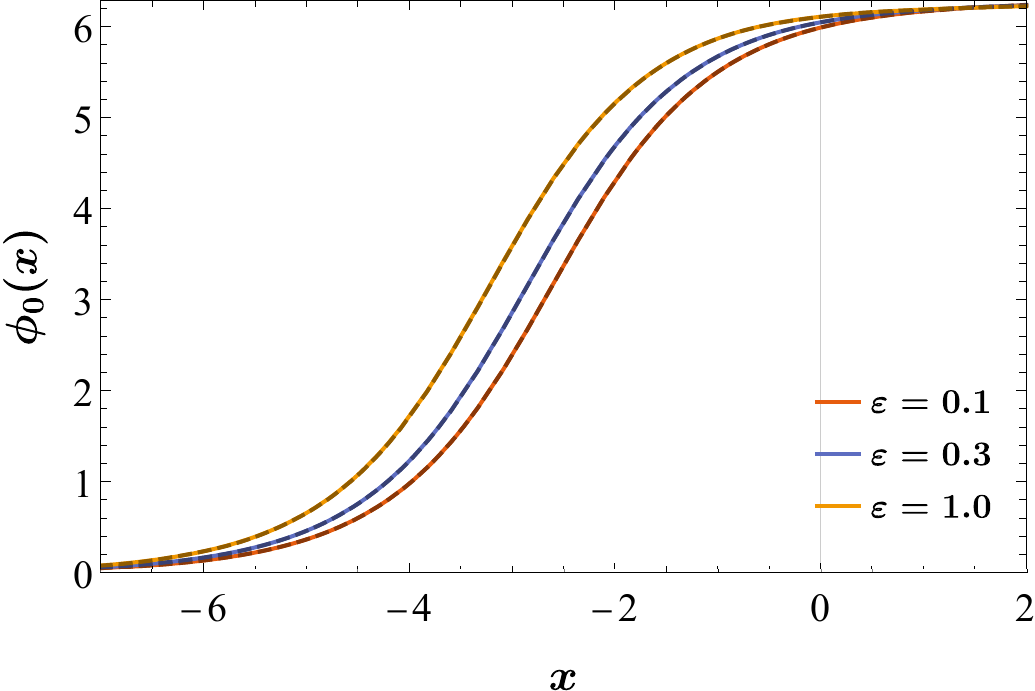}}}
    \caption{On the left panel, the $\chi(x)$ value determined from Eq. \eqref{linear-2+} for different values of $\varepsilon$ is shown. On the right
    panel, the solid line shows the sum of the kink ansatz $\phi_K(x)$ and $\chi(x)$ value for different $\varepsilon$, while the dashed line is the corresponding static solution  determined from original field model. In each case $\alpha=0.01$, while bias current is equal to $0.0045$, $0.0093$, and $0.017$ respectively for $\varepsilon = 0.1, 0.3,$ and $1.0$. }
    \label{fig_04}
\end{figure}

The stability of this configuration is tested in the standard way i.e., we linearize according to: $\phi(t,x) = \phi_0(x) + \psi(t,x).$ The equilibrium configuration itself can be roughly described, as before, by the sum of the free kink and the deformation $\phi_0(x) = \phi_K(x)+\chi(x)$, with the latter now being characterized in addition to the inhomogeneity, also by the bias current. The deformation satisfies the equation
\begin{equation}
\label{linear-2+}
    -\partial_x \left( {\cal F}(x)\partial_x \chi \right) + ( \cos \phi_K ) \chi = \varepsilon\partial_x\left( g(x)\partial \phi_K\right) - \Gamma.
\end{equation}
The left figure \ref{fig_04} shows the form of the correction describing the kink deformation coming from inhomogeneities. The right figure once again compares the static configuration obtained from the field model \eqref{sine-gordon-static} and the configuration obtained as the sum of the kink solution of the homogeneous model (for $\varepsilon=0$) and the correction derived from the inhomogeneity $\chi$. The very good agreement between the two results for different values of $\varepsilon$ shows that the splitting of the configuration $\phi_0$ into the correction $\chi$ and the kink of the free model $\phi_K$ provides an accurate description of the static configuration.

On the other hand, to explore the state's spectral stability, we use the linearization decomposition $\psi(t,x) = e^{i \omega t} v(x)$ which, in turn, leads to the eigenvalue problem:
\begin{equation}
\label{linear-4+} \hat{{\cal L}} \, v(x) = - \partial_x \left( {\cal F}(x)\, \partial_x v(x)\right) + ( \cos \phi_0 ) \, v(x) = \lambda v(x) \, .
\end{equation}
The quantity $\lambda$ appearing in this equation is related to the eigenfrequency $\omega$ as follows $\lambda=\omega (\omega - i \alpha),$ and therefore
\begin{equation}
    \psi(t,x) = e^{-\frac{1}{2} \alpha t} e^{\pm i \Omega t} v(x) ,
\end{equation}
where $\Omega=\sqrt{\lambda - \frac{\alpha^2}{4}}.$ As long as the condition $\lambda>\alpha^2/4$ is satisfied, then one can observe damped oscillations around $\phi_0(x)$, i.e., the relevant fixed point is a stable spiral. On the other hand when $\alpha^2/4 \geq \lambda>0$ one can observe overdamped behaviour of the perturbations
\begin{equation}
    \psi(t,x) = e^{-\frac{1}{2} \alpha t} e^{\pm  \kappa t} v(x) ,
\end{equation}
where $\kappa=\sqrt{ \frac{\alpha^2}{4}-\lambda}.$ In this case, the relevant fixed point corresponds to a stable node. As we will see below, in this damped-driven case, the system does possess a stable attractor, however, in reconciling with the Hamiltonian picture above, this is not the sole stationary state of the system. Indeed, the former saddle point of the Hamiltonian case, typically breaks up (in the presence of damping and driving) through a saddle-node bifurcation into a persistent unstable configuration and an emergent stable one (per the above discussion). We will iterate on this point further through our effective description in what follows.

\begin{figure}[ht]
    \centering
    \subfloat{{\includegraphics[height=5cm]{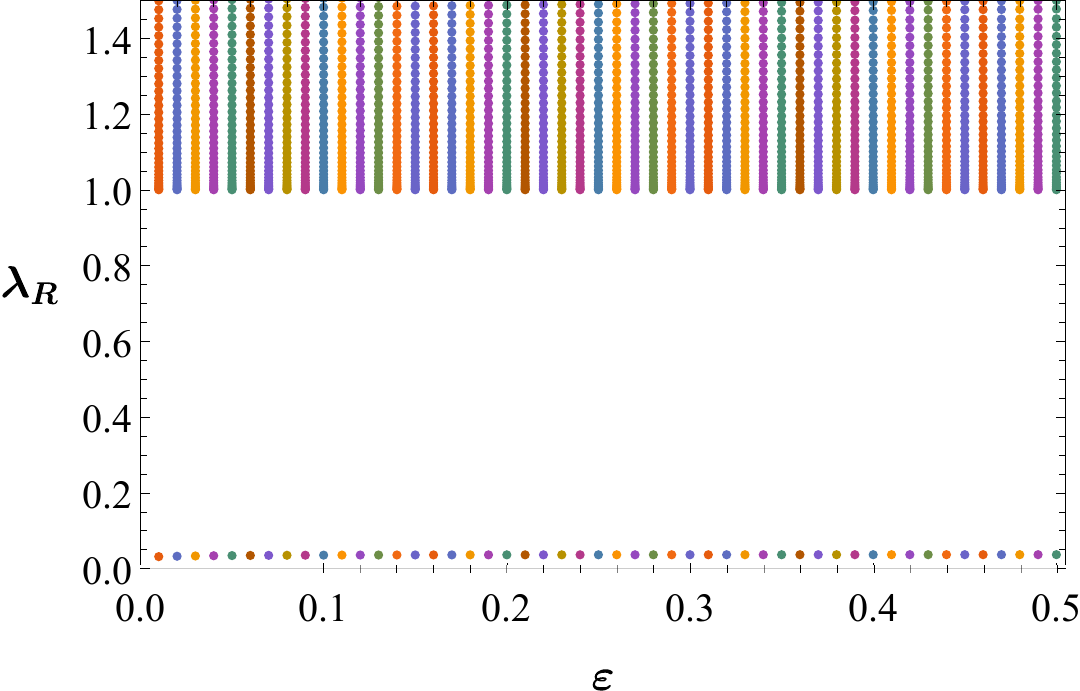}}}
    \quad
    \subfloat{{\includegraphics[height=5.1cm]{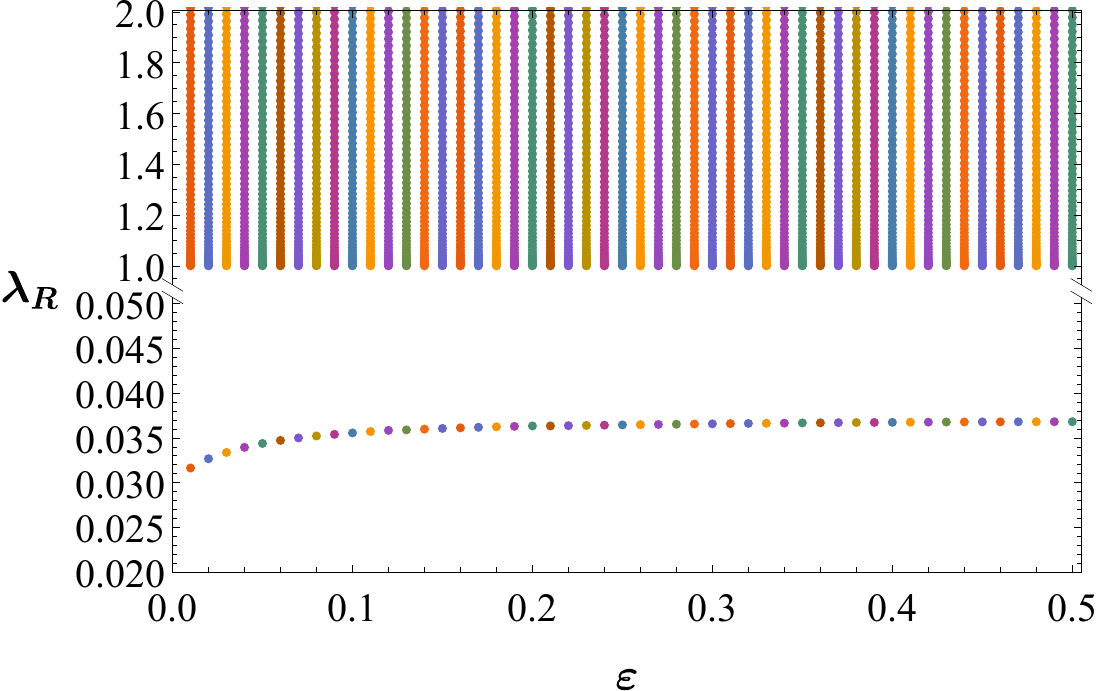}}}
    \caption{ Dependence of $\lambda_R=\omega^2$ on $\varepsilon$ for the static kink solution in the damped-driven sG model, where $\lambda_R$ is the real part of $\lambda$ (see under Eq. \eqref{linear-4+}). The graph regards the case of $\alpha=0.01$ and $\Gamma=0.001$ for each $\varepsilon$.}
    \label{fig_05}
\end{figure}

\section{Effective description of kink - inhomogenity interaction}

\subsection{Approximations based on one degree of freedom}
Having described the statics of the kink in the presence of the inhomogeneity, we now turn to the corresponding model dynamics in what follows. In this subsection, we will obtain an approximate description of the field system based on three approximation methods. The first method is based on a conservative Lagrangian of the system (i.e., an effectively variational method). The second method involves projecting the field equation onto the zero mode of the kink solution (the so-called translational mode, associated with the relevant invariance in the homogeneous limit), while the third one is based on a non-conservative Lagrangian. The last two methods allow the construction of an effective model even if there is a dissipation in the field equation, while the first is solely applicable in the realm of conservative (Lagrangian/Hamiltonian) systems.

\subsubsection{The construction based on a conservative lagrangian}
We start from the variation of the Lagrangian density 
\begin{equation}
\label{lanD_fsg}
\mathcal{L}_{FSG} =\mathcal{L}_{SG} + \mathcal{L}_{\varepsilon} = \mathcal{L}_{SG} - \frac{1}{2}\varepsilon g(x)(\partial_x\phi)^2,
\end{equation}
where $\mathcal{L}_{\varepsilon}$ describes the inhomogeneity present in the system and ${\cal L}_{SG}$ is the Lagrangian density of the (unperturbed) 
sine-Gordon model
\begin{equation}
\label{lanD_sg}
\mathcal{L}_{SG} = \frac{1}{2}(\partial_t\phi)^2 - \frac{1}{2}(\partial_x\phi)^2 - (1-\cos\phi).
\end{equation}
A recent discussion of such variational methods for the sG model, including in higher-dimensional settings can be found, e.g., in~\cite{ricardo}.

\begin{figure}[ht]
    \centering
    \subfloat{{\includegraphics[width=7.35cm]{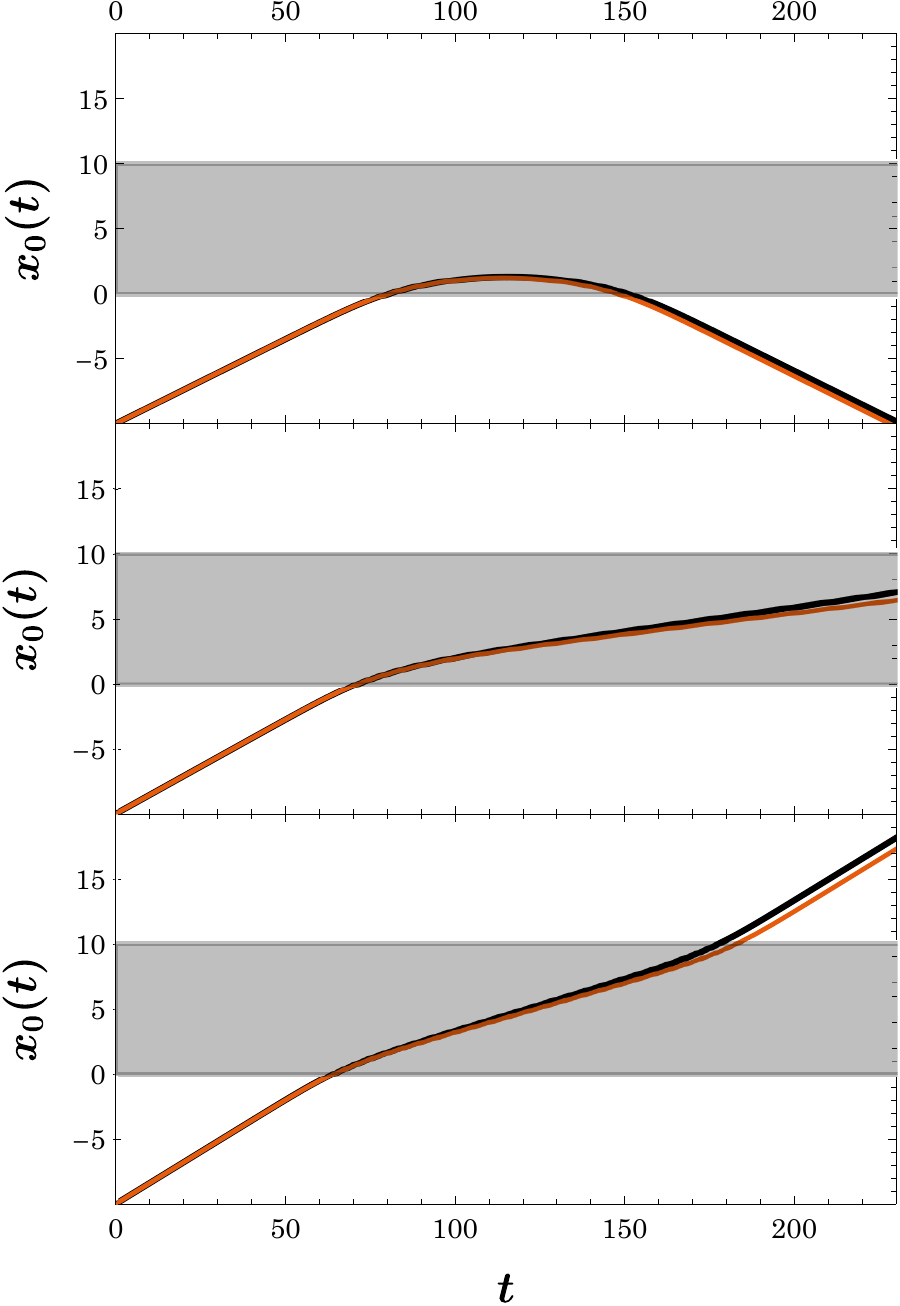}}}
    \quad
    \subfloat{{\includegraphics[width=7.5cm]{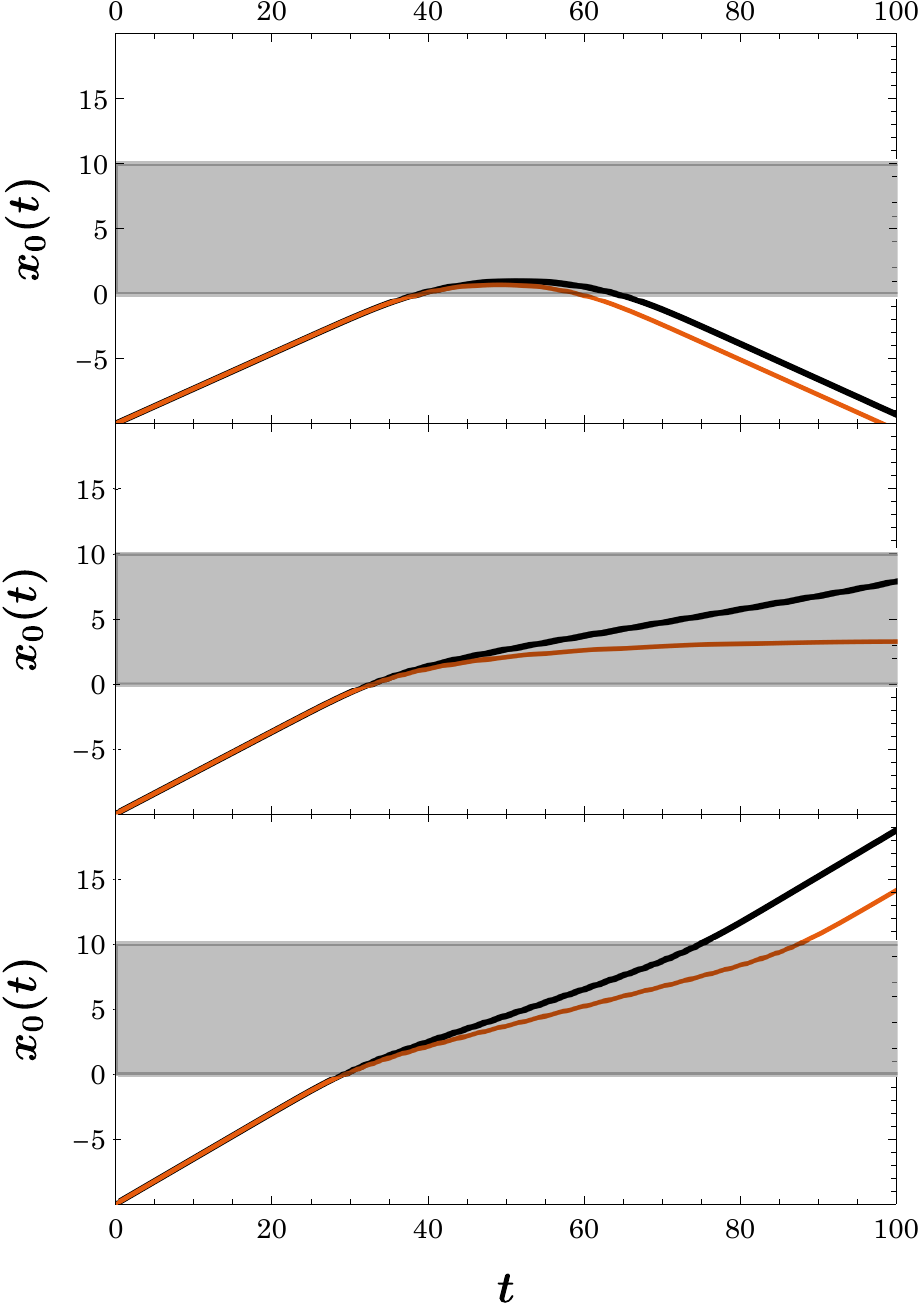}}}
    \caption{Comparison of the position of the center of mass of the kink for the solution from the original field model (black line) and the model with one degree of freedom (red line). The figures in the left panel are prepared for $\varepsilon = 0.01$ and velocities (starting from the top) $0.13$, $0.145$ and $0.16$. In the right panel $\varepsilon = 0.05$ and velocities from the top are $0.27$, $0.315$ and $0.35$.}
    \label{fig_06}
\end{figure}

\begin{figure}[ht]
    \centering
    \subfloat{{\includegraphics[width=7.45cm]{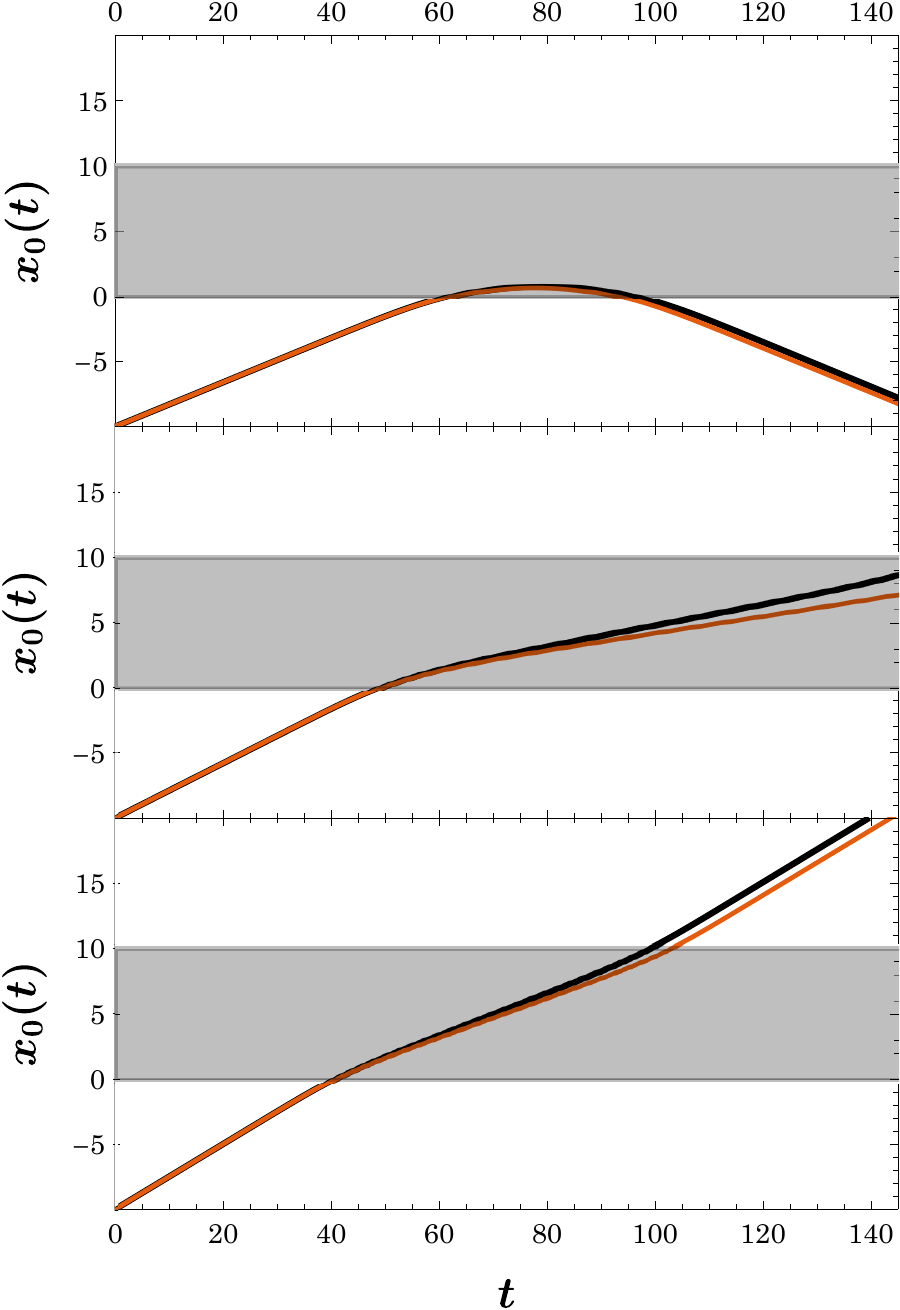}}}
    \quad
    \subfloat{{\includegraphics[width=7.7cm]{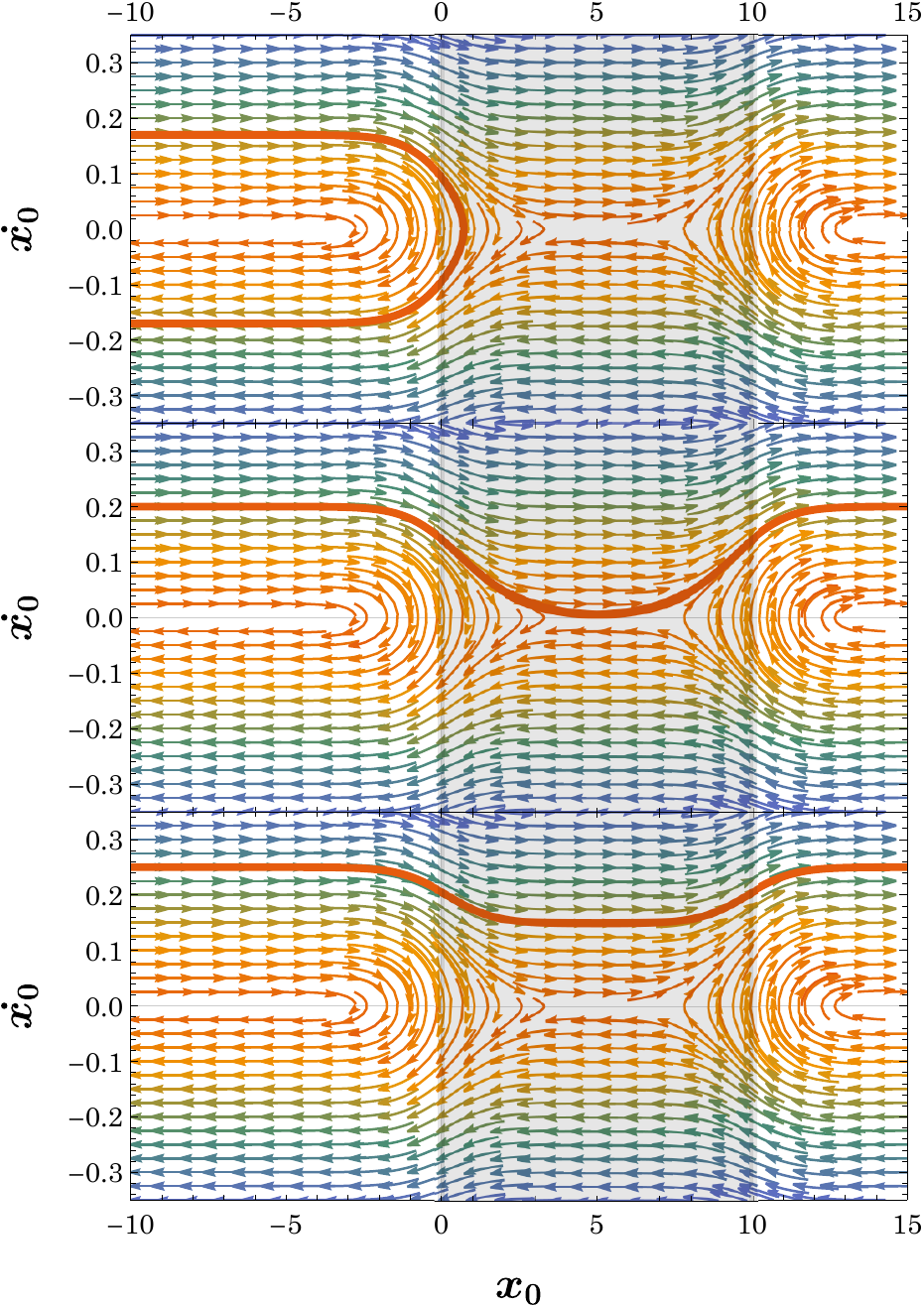}}}
    \caption{Comparison of the position of the center of mass of the kink for the solution from the original field  model (black line) and the model with one degree of freedom (red line). In the figures on the left $\varepsilon = 0.02$ and velocities from top $u=0.17$, $u=0.21$, and $u=0.25$. On the right are the phase diagrams corresponding to the same parameter values. The gray area represents the position of the inhomogeneity.}
    \label{fig_07}
\end{figure}

The reduction of the original PDE to a model with one degree of freedom is based on the use of the kink ansatz with the position of the kink used as a collective variable: 
\begin{equation}
\label{an_1dof}
\phi_{k}(t,x) = 4\arctan e^{x-x_0(t)}.
\end{equation}
By inserting the kink ansatz into the Lagrangian density \eqref{lanD_fsg} and then integrating with respect to the spatial variable, we obtain the effective Lagrangian for the variable $x_0(t)$
\begin{equation}
\label{lan_fsg}
L_{FSG} = \int_{-\infty}^{+\infty}dx \,{\cal L}_{FSG} = L_{SG} + L_{\varepsilon}=L_{SG} - \frac{1}{2}\varepsilon \int_{-\infty}^{+\infty}dx g(x)(\partial_x\phi_K)^2.
\end{equation}
The form of the interaction is uniquely determined by the function $g(x).$ In the case where $g(x)=0$ we obtain the Lagrangian of the free particle (after rescaling by the multiplicative constant and eliminating the additive one)
\begin{equation}
\label{lan_sg_val}
L_{FSG}=L_{SG} = \frac{1}{2}\Dot{x}_0^2.
\end{equation}
If $g$ is a nontrivial position-dependent function, then the effective Lagrangian is enriched by a potential energy landscape describing the interaction of the kink with the existing inhomogeneity. For example, if the function $g$ consists of unit step functions $g(x)=\theta(x)-\theta(x-L) $ then the Lagrangian assumes the form
\begin{equation}
\label{lan_full}
L_{FSG} = \frac{1}{2}\Dot{x}_0^2 -\varepsilon (\tanh(x_0)-\tanh(x_0-L) ).
\end{equation}
The equation of motion in this case is  the following: 
\begin{equation}
\label{1dof_eom}
\Ddot{x}_0 = -\varepsilon (\sech^2(x_0)-\sech^2(x_0-L)).
\end{equation}
On the other hand, in the previous section we used the function $g(x) = \tanh(x)-\tanh(x-L)$.  In this case the effective Lagrangian reads:
\begin{equation}    
\label{lan_full+}
L_{FSG} = \frac{1}{2}\Dot{x}_0^2 - \frac{1}{2}\varepsilon \left\{ \coth(x_0) - \frac{x_0} {\sinh^2 (x_0)}  - \left( \coth(x_0-L) - \frac{x_0-L} {\sinh^2 (x_0-L)} \right) \right\}.
\end{equation}
The potential energy landscape is provided by the term after the $(-)$ sign in Eq.~(\ref{lan_full+}) (or similarly in Eq.~(\ref{lan_full})) and clearly illustrates the existence of a local maximum corresponding to the saddle static kink configuration. The equation of motion for the collective variable is
\begin{equation}
\begin{multlined}
\label{1dof_eom_full}
\Ddot{x}_0 =  \varepsilon \left(
 \frac{1 - x_0 \coth{x_0}}{\sinh^2{x_0}} - \frac{1 - (x_0-L) \coth{(x_0-L)}}{\sinh^2{(x_0-L)}}
\right).
\end{multlined}
\end{equation}
The trajectories obtained from the last equation are compared with center-of-mass trajectories following from the field equation (Eq.~\eqref{sine-gordon} with $\alpha=0$ and $\Gamma=0$). Figure \ref{fig_06} shows a good agreement between the effective model and the full field model for small values of $\varepsilon$. The left panel in this figure corresponds to $\varepsilon=0.01$ while the right panel contains results for $\varepsilon=0.05$. Each of the panels consists of three figures. The top figure describes the kink reflecting from the inhomogeneity. The initial speed in this case $u=0.13$ is lower than the critical velocity. The second figure in this panel represents the interaction of the kink whose initial speed $u=0.145$ is close to the critical velocity. The bottom figure demonstrates the kink passing over the barrier for initial speed $u=0.16$ exceeding the critical value. Similarly, in the figures of the right panel, the velocities are smaller $u=0.27$, close to the critical value $u=0.315$ and above $u=0.35$ the critical speed. The critical velocities for which the agreement takes place are approximately limited to $0.25$. For larger values of this parameter, inconsistencies become more significant.
The trajectories corresponding to $\varepsilon=0.02$ are presented in Figure \ref{fig_07}. The initial velocities of the kink are, starting from the top, $0.17$, $0.21$ and $0.25$. The right panel shows the course of these trajectories (represented by the red line) on the background of the phase space.
The phase diagrams show an unstable fixed point at the center of the barrier. 

\begin{figure}[ht]
    \centering
    \includegraphics[height=6cm]{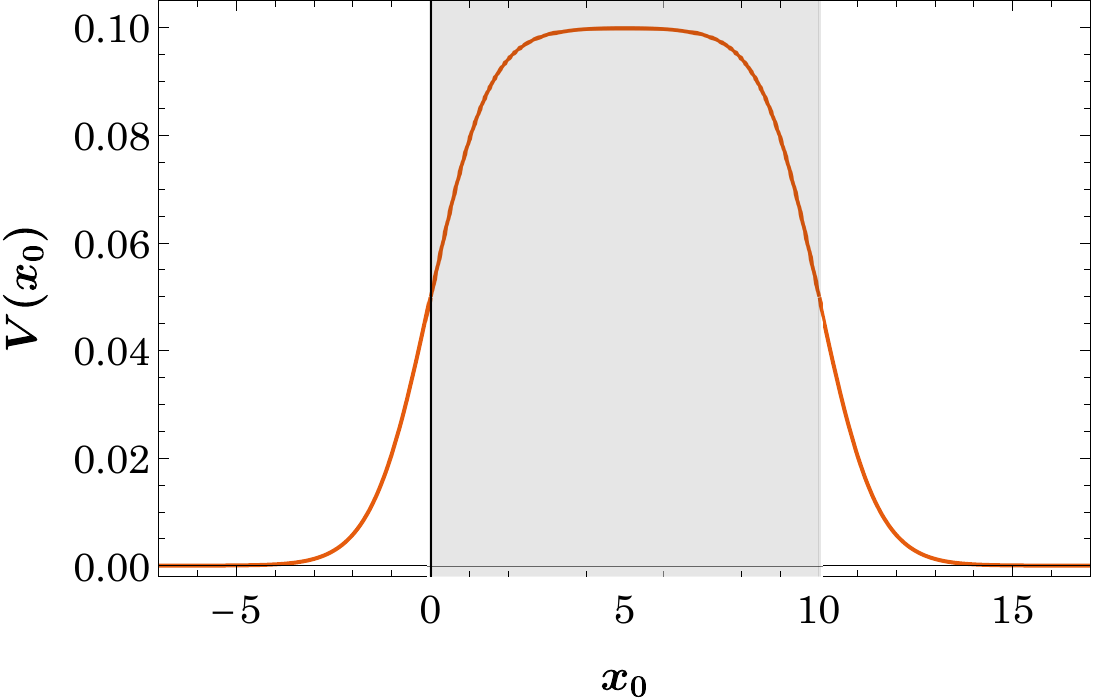}
    \caption{Graphical representation of the potential represented by the second term (preceded by a minus sign) of eq. \eqref{lan_full+}. In the figure, we assumed 
    $\varepsilon = 0.1$ and $L=10$.}
    \label{fig_08}
\end{figure}

\begin{figure}[ht]
    \centering
    \includegraphics[height=6cm]{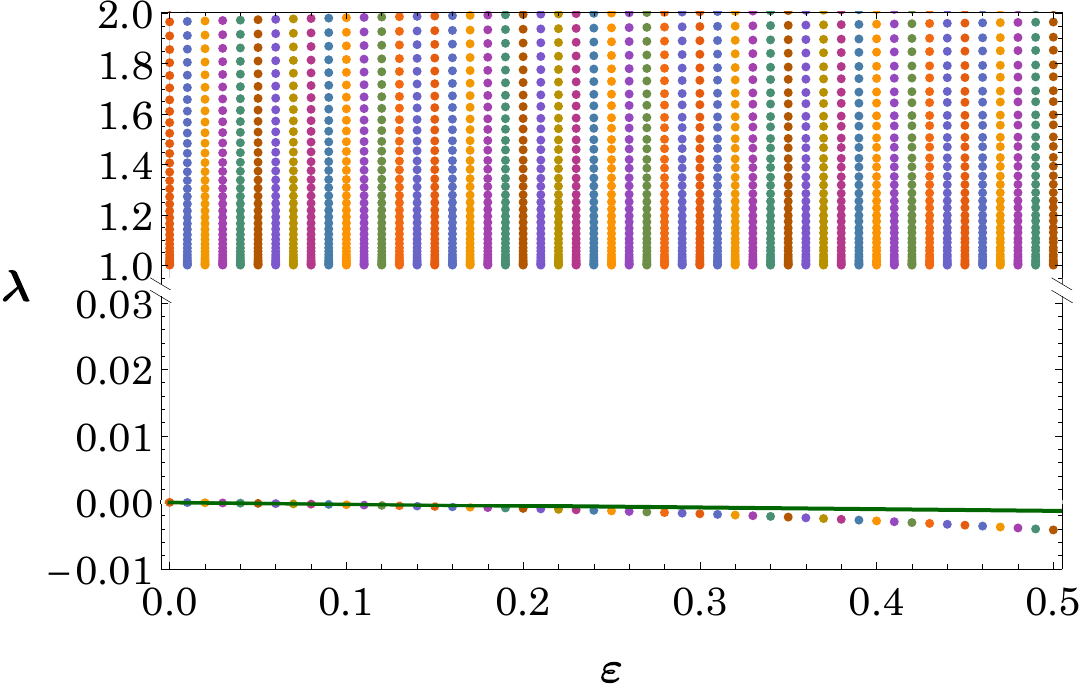}
    \caption{Comparison of the squared eigenfrequencies
    $\lambda=\omega^2$ obtained from the field model (in case of $\alpha=0$, $\Gamma=0$) with the ones determined from the effective model with one degree of freedom (green line).}
    \label{fig_09}
\end{figure}

The corresponding potential energy landscape representing the relevant energy maximum can be seen in Fig.~\ref{fig_08}. In the case we are considering, the location of the fixed point is $x_0=5.$
Regarding the linear stability in the effective particle model, only one (unstable mode) is naturally present, pertaining to the formerly translational mode
of the homogenous sG. A comparison of this mode with the spectrum of linear excitations of the field model ($\alpha=0$, $\Gamma=0$) can be found in Figure \ref{fig_09}. The green line represents the result obtained from the model with one degree of freedom, while the points represent the spectrum of the $\hat{{\cal L}}$ operator. Quantitative agreement occurs only for small values of the $\varepsilon$ parameter, yet the qualitative agreement between the two is
clearly evident.
\subsubsection{The method of projecting onto the zero mode} 
\begin{figure}[ht]
    \centering
    \subfloat{{\includegraphics[width=7.45cm]{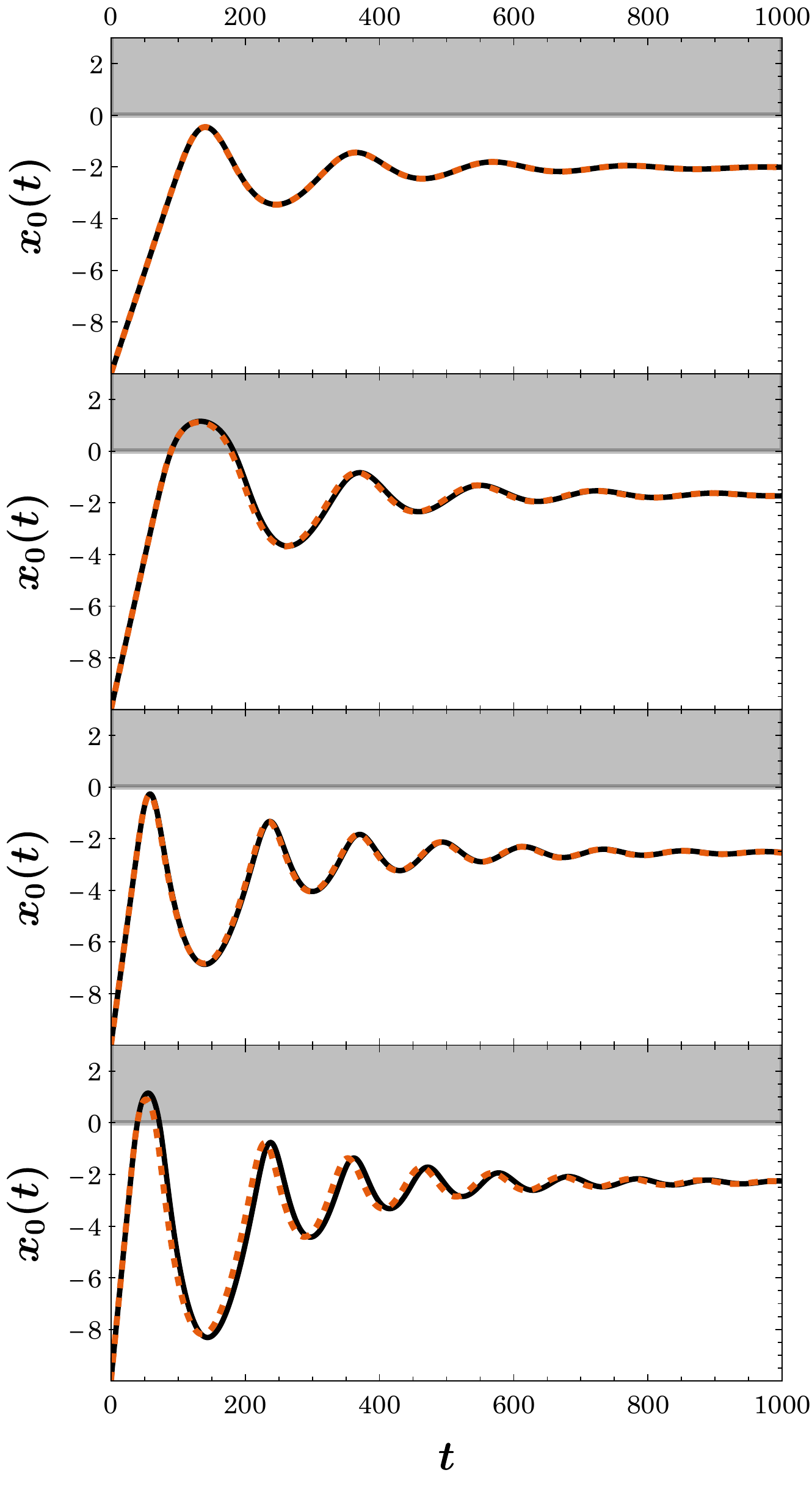}}}
    \quad
    \subfloat{{\includegraphics[width=7.5cm]{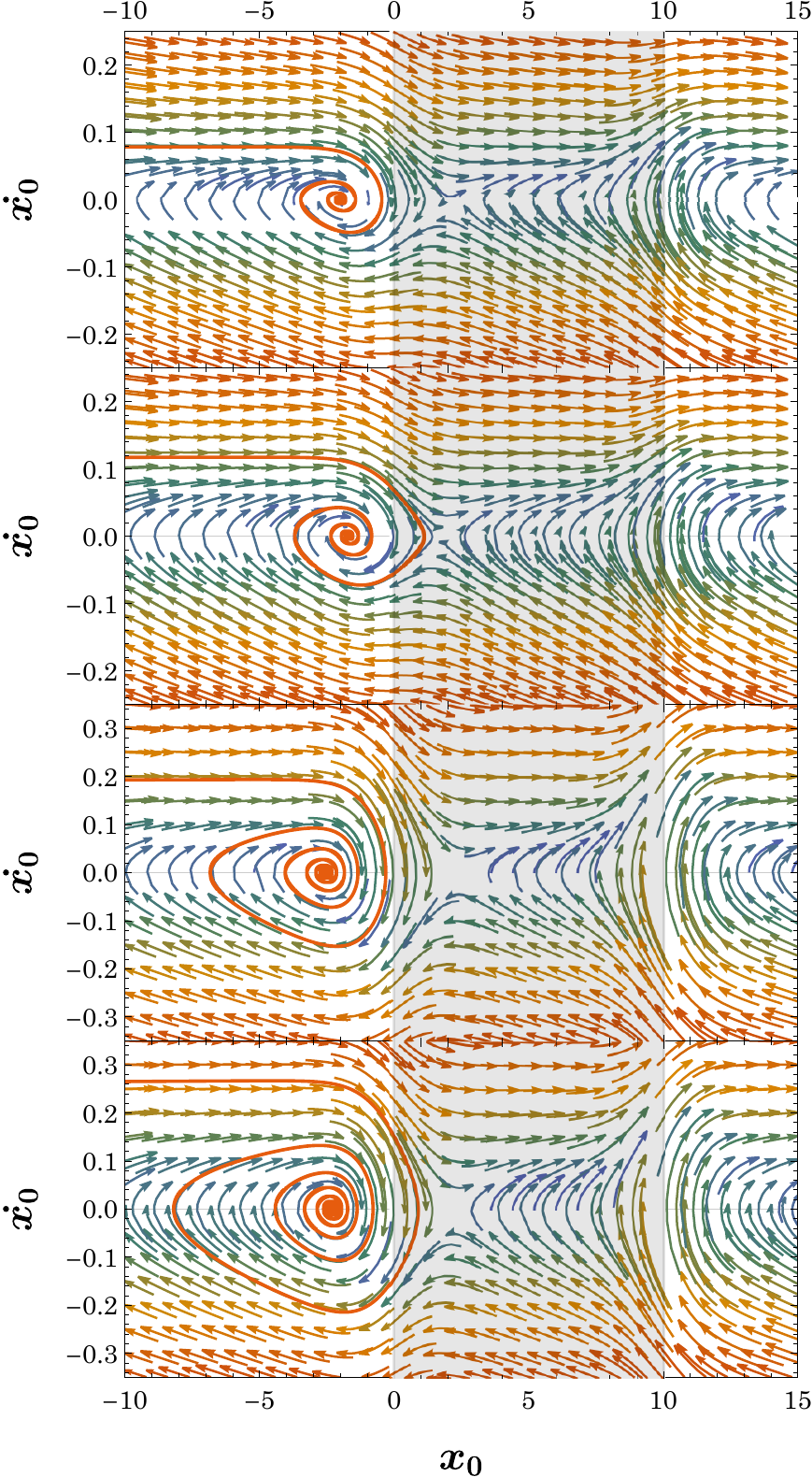}}}
    \caption{Comparison of the position of the center of mass of the kink for the solution from the original field  model (black line) and the model with one degree of freedom (red dashed line). In the figures on the left $\varepsilon = 0.01$ and bias current from the top $0.001$ (top row), $0.0015$ (second row), then $\varepsilon = 0.05$ and bias current $0.0025$ (third row), $0.0035$ (fourth row). In each case, the dissipation is equal to $0.01$. On the right are the phase diagrams corresponding to the same parameter values. The gray area represents the position of the inhomogeneity.}
    \label{fig_10}
\end{figure}

\begin{figure}[ht]
    \centering
    \includegraphics[height=6cm]{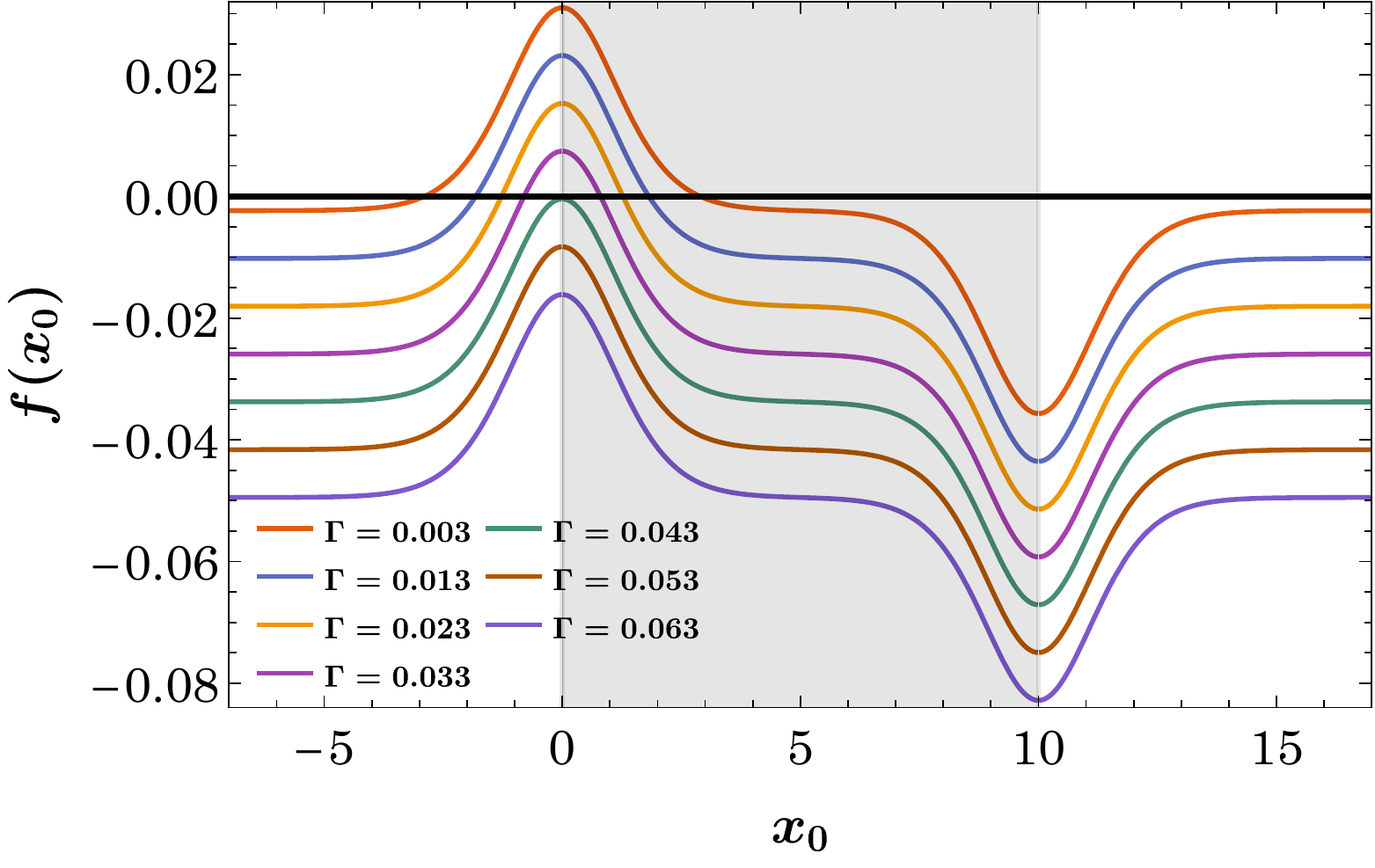}
    \caption{Graphical representation of the position (and existence) of fixed
points for different values of bias current (see equation
\eqref{model-2fx0}). 
These are the places where the function $f(x_0)$ vanishes.
In all cases $\varepsilon = 0.1$.}
    \label{fig_11}
\end{figure}

In addition to the above method of effective theory construction, other approaches are used in the literature. One of them is the zero mode projection method; see, e.g., a relevant discussion in~\cite{spiegel}. This method, unlike the standard method based on the conservative Lagrangian, is based on the PDE itself and does not hinge on the variational structure of the problem. As such, it allows for an effective description of systems containing dissipative terms. With this in mind, we can construct an effective model for the sine-Gordon model with dissipation described by the equation \eqref{sine-gordon} with $\alpha \neq 0$ and $\Gamma \neq 0$. Practically, we insert the kink ansatz
\begin{equation}\label{model-2b}
\phi(t,x)=4 \arctan e^{\xi(t,x)},
\end{equation}
into the field equation \eqref{sine-gordon}. This substitution results in the equation
\begin{equation}\label{model-2c}
   \left(\ddot{\xi}-\xi'' + \alpha \dot{\xi}\right) \partial_{\xi} \phi
+\left(1+ \dot{\xi}^2-{\xi'}^2 \right)\, \partial_{\xi}^2 \phi =
\varepsilon (\partial_x \,g) \, \xi' \,\partial_{\xi} \phi  +
\varepsilon g \,  \left( \xi'' \partial_{\xi} \phi + {\xi'}^2
\partial_{\xi}^2 \phi \right) - \Gamma \, .
\end{equation}
The dot denotes the derivative with respect to the time variable while the prime denotes the derivative with respect to the spatial variable. If we want to obtain a model describing the dynamics of one collective variable, i.e., the variable that determines the position of the kink, we take a particular form of the function
$\xi = \xi(t,x)$ i.e.
\begin{equation*}
    \xi (t,x) = x -x_0(t) .
\end{equation*}
With this substitution, the \eqref{model-2c} equation is reduced to a much simpler form
\begin{equation}\label{model-2d}
    \left(- \ddot{x}_0 - \alpha \dot{x}_0 \right) \partial_{\xi} \phi
+ \dot{x}_0^2 \, \partial_{\xi}^2 \phi - \varepsilon (\partial_x
\,g) \,\partial_{\xi} \phi  - \varepsilon g \,
 \partial_{\xi}^2 \phi  + \Gamma =0 .
\end{equation}
The final step is the projection of the above equation onto the zero mode, which consists of integration with the kink ansatz derivative representing the zero mode
of the (former) homogeneous model, i.e.,
\begin{equation}\label{model-2e}
 \int_{-\infty}^{+\infty} d \xi \left( \left(- \ddot{x}_0 - \alpha \dot{x}_0 \right) \partial_{\xi} \phi
+ \dot{x}_0^2 \, \partial_{\xi}^2 \phi - \varepsilon (\partial_x
\,g) \,\partial_{\xi} \phi  - \varepsilon g \,
 \partial_{\xi}^2 \phi  + \Gamma \right) \partial_{\xi} \phi =0 .
\end{equation}
When performing integration, we must adopt a particular form of the function $g$ describing the inhomogeneity. In this part, we assume similarly to the previous one $g(x)=\tanh x - \tanh(x-L)$. As a result of the integration, we obtain an effective equation of the form
\begin{equation}\label{model-2f}
    \ddot{x}_0 + \alpha \dot{x}_0 + \varepsilon \left( \frac{x_0 \coth x_0 - 1}{\sinh^2x_0} - \frac{(x_0-L) \coth (x_0-L) - 1}{\sinh^2(x_0-L)}\right) =
    \frac{\pi}{4} \Gamma
\end{equation}
Note that after removing the terms containing the coefficients $\alpha$ and $\Gamma$, the above equation reduces to equation \eqref{1dof_eom_full} and therefore, in this case, the numerical results are included in Figures \ref{fig_06} and \ref{fig_07}. On the other hand, if we have dissipation and current present in the system, then below the critical velocity, we observe the effect of multiple reflections from an inhomogeneity. The course of this process is shown in Figure \ref{fig_10}. In these figures, we assume that the dissipation coefficient is $\alpha=0.01$ and that the bias current $\Gamma$ is equal to $0.001$, in the top left figure and  $0.0015$, in the second 
row of the figure, respectively. In both figures, $\varepsilon=0.01$. The initial velocity of the kink in all cases with dissipation is chosen to be equal to the stationary velocity obtained in the article \cite{Scott1978}, i.e. according to the formula:
\begin{equation}
\label{stationary_velocity}
u_{s}=\frac{1}{\sqrt{1+\left(\frac{4\alpha}{\pi\Gamma}\right)^2}}.
\end{equation}
Equation \eqref{sine-gordon} at $\varepsilon=0$ has a solution in the form of a kink moving with constant velocity only when the dissipation occurring in the system is exactly balanced by the forcing in the form of a bias current i.e. only for stationary velocity $u_s$. The initial condition describing a kink with velocity $u>u_s$ always, due to the existence of dissipation, slows down to a value of $u_s$. On the other hand, the initial condition with $u<u_s$, as a result of forcing, accelerates to $u_s$, i.e., to the velocity at which there is a balancing of forcing with dissipation. For this reason, if the initial velocity takes the value $u_s$ (for $\varepsilon \neq 0$) the velocity changes are related only to the interaction with the inhomogeneity. Note that in both of these cases the kink trajectory resulting from the field model (solid black line) coincides with the trajectory obtained from the effective model (dashed red line). As one can see, in these runs the kink that has too low velocity to penetrate the barrier hence bounces back; yet, the presence of constant forcing causes successive returns toward the barrier. Due to the existence of dissipation in the system, the amplitude of subsequent reflections is reduced. From a dynamical systems perspective, this clearly suggests the existence of a fixed point in the form of a stable spiral, which is unveiled in the phase portrait illustrated in the right panel of the figure. The third and fourth rows of the figure illustrate the relevant features for a larger value of the parameter $\varepsilon=0.05$ characterizing the inhomogeneity. Respectively, the corresponding bias currents are $\Gamma=0.0025$, and $\Gamma=0.0035$. It can be seen that in these cases the deviations of the continuous black and dashed red lines are insignificant and therefore the right panel contains only the phase space of the effective model. In the phase space presented in the figures of the right panel, one can identify two fixed points. The point located on the left side of the barrier, as discussed above, represents a stable spiral. The trajectories shown in the figures of the left panel are represented on the right panel by using red spirals. The second fixed point is located in the barrier area (representing the analogue/remnant of the fixed point present in the conservative case) and has the character of a saddle point. In all panels of Figure~\ref{fig_10}, the gray area represents the position of inhomogeneity, or more precisely the area located between $x=0$ and $x=L$. The stable manifold of the saddle represents on each side the separatrix between the trajectories that are transmitted and those that are reflected.
The location of the fixed points can be determined by referring to the equation \eqref{model-2f}. By transferring the bias current to the left side and zeroing out all derivatives with respect to time to identify the fixed points, we obtain a function $f(x_0)$
\begin{equation}\label{model-2fx0}
    f(x_0) \equiv \varepsilon \left( \frac{x_0 \coth x_0 - 1}{\sinh^2x_0}
    - \frac{(x_0-L) \coth (x_0-L) - 1}{\sinh^2(x_0-L)}\right) -
    \frac{\pi}{4} \Gamma =0 ,
\end{equation}
whose zeros indicate the desired equilibria. Figure \ref{fig_11} shows the positions of these points depending on the value of the current. It can be seen, that for currents below the threshold value, i.e., for $0.003$, $0.013$, $0.023$ and $0.033$ there are two fixed points, a stable one on the left and an unstable one on the right. Physically, the presence of a stable fixed point is related to the fact that the kink, not having enough energy to cross the barrier bounces off it before getting trapped
on the stable spiral fixed point. On the other hand, the constant forcing presses the kink to move towards the barrier. At the same time, the kink loses energy due to dissipation which leads to its eventual stopping. The presence of the second fixed point can be interpreted as the kink sliding off the barrier, and is an effective remnant of the conservative
case with $\Gamma=0$.
For the threshold value of the bias current, i.e., $\Gamma=0.043$ only an unstable point remains. When the value of the current exceeds the threshold value (e.g., for $\Gamma=0.053$ and $\Gamma=0.063$) the fixed points do not occur. In these cases, the barrier is unable to stop the kink because the energy provided by the drive is too high. 
For negative values of the current, the situation is symmetrical with respect to the barrier, as long as the kink moves from the right side toward the left, i.e., the stable point is on the right side of the barrier while the unstable one is located on the barrier.

Similar to Figure \ref{fig_06}, Figure \ref{fig_12} shows the kink interaction with the inhomogeneity for two values of $\varepsilon$. The left panel corresponds to $\varepsilon=0.01$ while the right one corresponds to $\varepsilon=0.05$. Initial speeds are determined by the formula \eqref{stationary_velocity}. In all plots, the dissipation coefficient is equal to $\alpha=0.01$. In the left panel the bias current $\Gamma$ is equal, respectively, in the 3 rows: $0.0015$, $0.00155$ and $0.0016$, while in the right panel $0036$, $0.00384$ and $0.0041$. In these two panels the upper figures show a reflection of the kink from the barrier (leading to its eventual trapping). The middle figures represent the interaction of the kink with velocity close to the critical speed, ultimately in these cases leading to transmission. The bottom figures show the passage of a kink over a barrier for speeds above the critical velocity. It can be seen that for $\varepsilon=0.01$ the agreement of the prediction of the approximate equation and the original one is very good while for $\varepsilon=0.05$ we observe nontrivial discrepancies, which suggest the potential of the latter scenario for further improvement, as concerns its theoretical description. Finally, it is worth noting that the equations of motion in the case of zero mode projection and the method based on the conservative Lagrangian are identical for $\alpha=0$ and $\Gamma=0$. This behavior for effective models with one degree of freedom is not a coincidence. In the appendix, we show that in the case without dissipation the two effective descriptions are equivalent.

\subsubsection{The method based on non-conservative Lagrangian}
\begin{figure}[ht]
    \centering
    \subfloat{{\includegraphics[width=7.5cm]{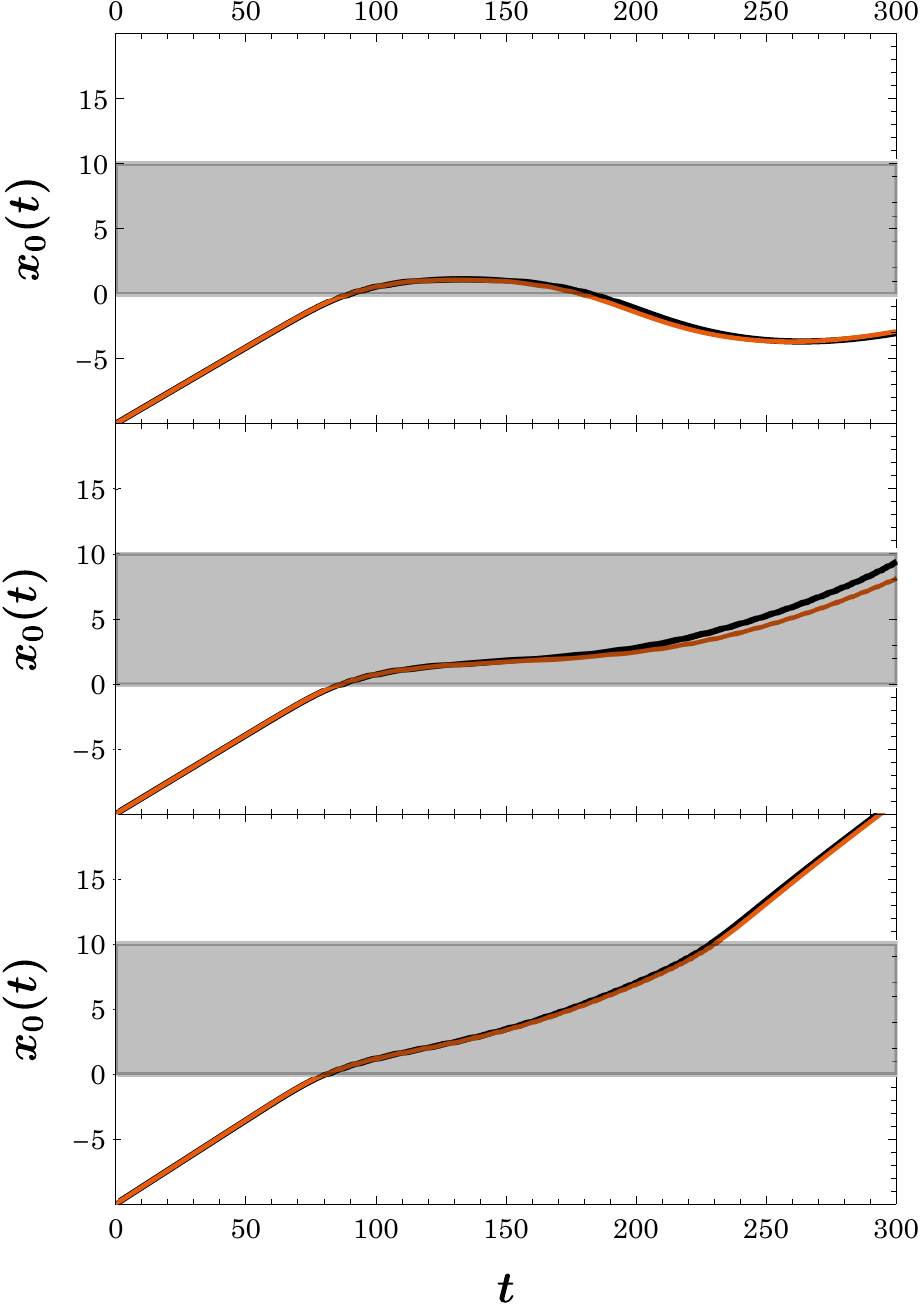}}}
    \quad
    \subfloat{{\includegraphics[width=7.35cm]{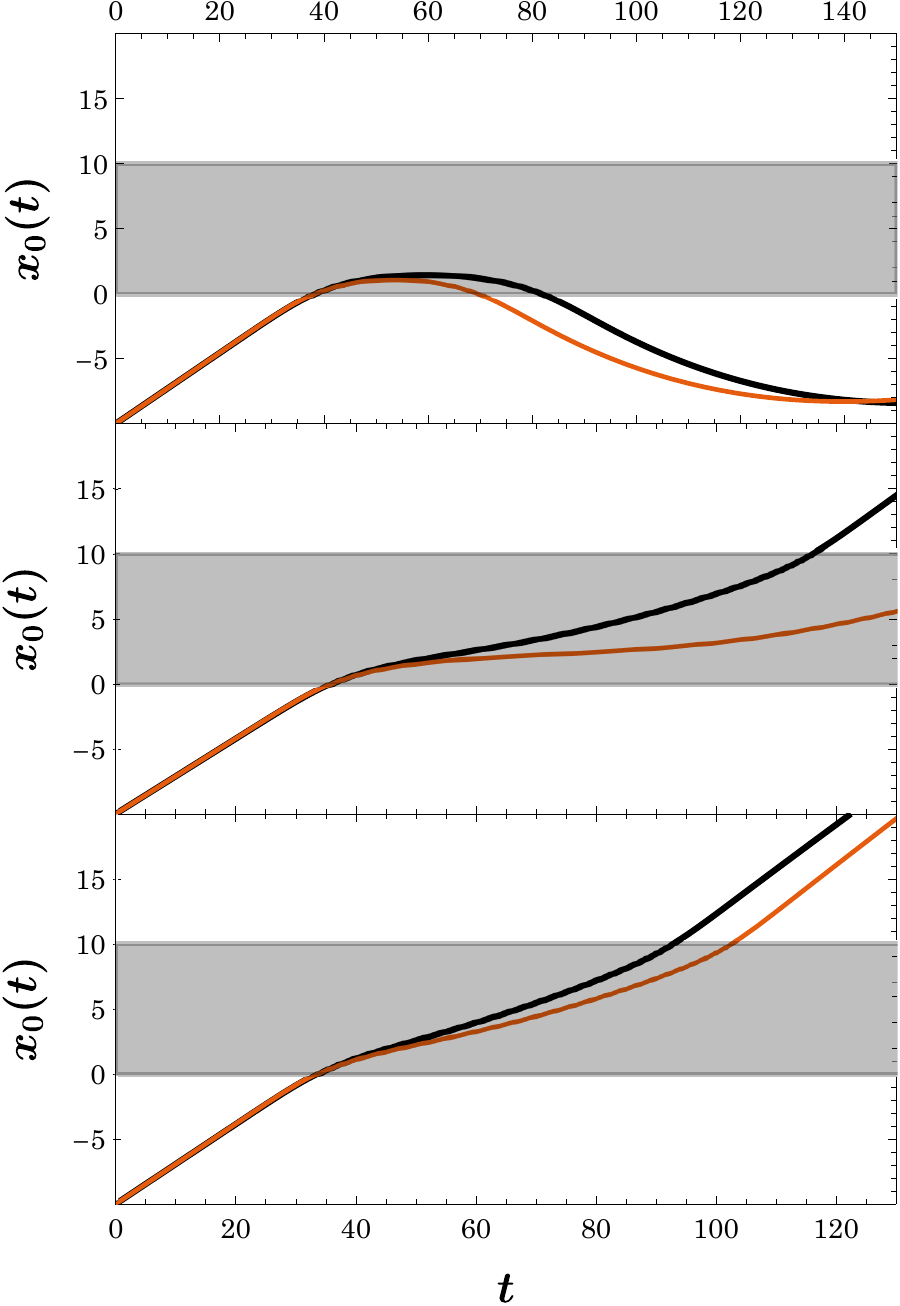}}}
    \caption{Comparison of the position of the center of mass of the kink for the solution from the original field model (black line) and the model with one degree of freedom (red line). The left panel contains figures for inhomogeneity $\varepsilon=0.01$ and bias current with values starting from the top $0.0015$, $0.00155$ and $0.00165$. On the right panel results for $\varepsilon=0.05$ are presented. Starting from the top the bias current is $0.0036$, $0.00384$ and $0.0041$. In each case, the disipation was equal to $0.01$.}
    \label{fig_12}
\end{figure}

Another proposal for obtaining both the original field equation (containing the dissipation) and the effective equation of motion for the collective coordinate reduced description is a method based on the non-conservative Lagrangian density \cite{Galley2013,Kevrekidis2014}. The field equation in this case can be obtained based on the standard conservative Lagrangian density $\mathcal{L}$ and the non-conservative contribution $\mathcal{R}$
\begin{equation}
\label{model-3a}
\partial_{\mu} \left( \frac{\partial \mathcal{L}}{\partial (\partial_{\mu}\phi)}\right)
- \frac{\partial \mathcal{L}}{\partial \phi} =
\lim_{\phi_{-}\rightarrow 0} \, \left\{ \lim_{ \phi_{+}\rightarrow
\phi } \left[ \frac{\partial \mathcal{R}}{\partial \phi_{-}} -
\partial_{\mu} \left( \frac{\partial \mathcal{R}}{\partial
(\partial_{\mu}\phi_{-})}\right)
 \right] \right\} .
\end{equation}
In the system we are currently considering, $\mathcal{L}=\mathcal{L}_{FSG}$ and $\mathcal{R}=-\alpha \phi_{-} \partial_t \phi_{+}-\Gamma \phi_{-}$. Here $\phi_{-}$ and $\phi_{+}$ are auxiliary fields with the property that in the so-called physical limit $\phi_{-} \rightarrow 0$ and $\phi_{+} \rightarrow \phi$. The equation \eqref{model-3a} written above reproduces the field equation \eqref{sine-gordon} with dissipation. The effective Lagrangian and the non-conservative potential at the effective level are obtained by inserting the kink ansatz into the densities and integrating over the spatial variable
\begin{equation}
\label{model-3b} L= \int_{-\infty}^{+\infty} \mathcal{L}
(\phi_K,\partial_{\mu} \phi_K) dx  , \, \, \, R=
\int_{-\infty}^{+\infty} \mathcal{R} (\phi_{K \pm},\partial_{\mu}
\phi_{K \pm}) dx .
\end{equation}
The effective equation has a similar structure to the original one
\begin{equation}
\label{model-3c} \partial_t \left( \frac{\partial L}{\partial
\dot{x}_0}\right)  - \frac{\partial L}{\partial x_0} =
\lim_{x_{-}\rightarrow 0} \, \left\{\lim_{ x_{+}\rightarrow x_0  }
\left[ \frac{\partial R}{\partial x_{-}} - \partial_t \left(
\frac{\partial R}{\partial \dot{x}_{-}}\right) \right] \right\} .
\end{equation}
After substituting the effective quantities into equation \eqref{model-3c}, we get
\begin{equation}\label{model-3d}
    \ddot{x}_0  + \varepsilon \left( \frac{x_0 \coth x_0 - 1}{\sinh^2x_0} - \frac{(x_0-L) \coth (x_0-L) - 1}{\sinh^2(x_0-L)}\right) = -\alpha \dot{x}_0
    + \frac{\pi}{4} \Gamma .
\end{equation}
Note that this equation is identical to equation \eqref{model-2f} and for $\alpha=0$ and $\Gamma=0$ reduces to equation \eqref{1dof_eom_full}. 
Per our previous discussion, the trajectories in this simplified case 
have been compared with the trajectories obtained in the original field model in Figure \ref{fig_06}, while the case with dissipation ($\alpha \neq 0$ and $\Gamma \neq 0$) has been shown in Figures \ref{fig_10} and \ref{fig_12}.
\subsection{Approximations based on two degrees of freedom}
We now turn to representations of the effective solitary wave dynamics using two degrees of freedom.
More specifically, we consider the position of the kink $x_0(t)$ and a parameter describing its effective inverse width parametrized
by $\gamma(t)$ through the ansatz:
\begin{equation}
\label{an_2dof}
\phi_{K}(t,x) = 4\arctan e^{\gamma(t)(x-x_0(t))}.
\end{equation}
Such a functional form is expected to allow us both to better describe the motional effects of the kink and also to capture the (potential) 
excitation of the kink's vibrational (internal breathing) mode.
We now proceed to provide the associated details, providing all
3 of the effective descriptions used before (one for the Hamiltonian
and two for the damped-driven problem).
\subsubsection{The construction based on a conservative Lagrangian (for $\alpha=0$ and $\Gamma=0$)} Similarly to the previous section, after integrating the model Lagrangian (\ref{lanD_fsg}) over the spatial variable $x$, we obtain in this case a sine-Gordon part of the effective Lagrangian of the form
\begin{equation}
\label{lan_sg_val_2dof}
L_{SG} =4\gamma\Dot{x}_0^2 +\frac{\pi^2}{3\gamma^3}\Dot{\gamma}^2-4\left(\gamma + \frac{1}{\gamma}\right).
\end{equation}
The form of the second part of the effective Lagrangian is strongly dependent on the shape of the inhomogeneity, i.e., on the analytical form of the function $g(x).$ For example, for a function consisting of a unit step  $g(x)=\theta(x)-\theta(x-L)$ this part of the Lagrangian is as follows
\begin{equation}
\label{lan_e_val_2dof}
L_{\varepsilon} = - 2\varepsilon\gamma \left[\tanh(\gamma x_0)  - \tanh(\gamma (x_0-L))\right] .
\end{equation}
Then, the equations of motion in this case (stemming from the effective Lagrangian $L=L_{SG}+L_{\varepsilon}$) have a relatively compact form
\begin{equation}
\begin{gathered}
\label{2dof_oem}
    \Ddot{x}+\frac{\Dot{\gamma}}{\gamma}\Dot{x}_0+\frac{1}{4}\varepsilon\gamma \left[\sech^2(\gamma x_0) -\sech^2(\gamma(x_0-L)) \right]=0, \\
    \frac{2 \pi^2}{3} \frac{\Ddot{\gamma}}{\gamma}  - \pi^2 \frac{\Dot{\gamma}^2}{\gamma^2}-4 \gamma^2 \Dot{x}^2+4\left(\gamma^2-1\right) + 2\varepsilon \gamma^2\left[\tanh(\gamma x_0)-\tanh(\gamma (x_0-L))\right]\\+ 2\varepsilon \gamma^3 \left[x_0 \sech^2(\gamma x_0)-(x_0-L) \sech^2(\gamma(x_0-L))\right]=0.
\end{gathered}
\end{equation}
Due to the high complexity and length of the formulas, we do not give the form of the equations obtained for $g(x)=\tanh(x)-\tanh(x-L)$, however, these equations are used when comparing the effective model with the field model \eqref{sine-gordon}. For completeness of description, the part $L_{\varepsilon}$ of the Lagrangian responsible for the interaction with the inhomogeneity for this case is given as:
\begin{dmath}
    L_{\varepsilon}=\frac{8 \gamma  \epsilon  e^{2 \gamma  x_0}}{\left(e^{2 \gamma  x_0}-1\right){}^2 \left(e^{2 \gamma  L}-e^{2 \gamma  x_0}\right){}^2} \left(e^{2 \gamma  L} \left(e^{2 \gamma  x_0}-1\right){}^2 \log \left(e^{2 \gamma  L}\right)+4 \sinh (\gamma  L) e^{2 \gamma  \left(L+x_0\right)} \left(-\cosh (\gamma  L)+\cosh \left(\gamma  \left(L-2 x_0\right)\right)+\log \left(e^{2 \gamma  x_0}\right) \sinh \left(\gamma  \left(L-2 x_0\right)\right)\right)\right) .
    \label{L-e}
\end{dmath}
As before, the effective Lagrangian $L_{FSG}$ consists of two parts, i.e., the effective Lagrangian of the free sine-Gordon model \eqref{lan_sg_val_2dof} and the lagrangian describing the interaction with inhomogeneity \eqref{L-e}. This effective Lagrangian is used to obtain the equations of motion.

\begin{figure}[ht]
    \centering
    \subfloat{{\includegraphics[width=7.4cm]{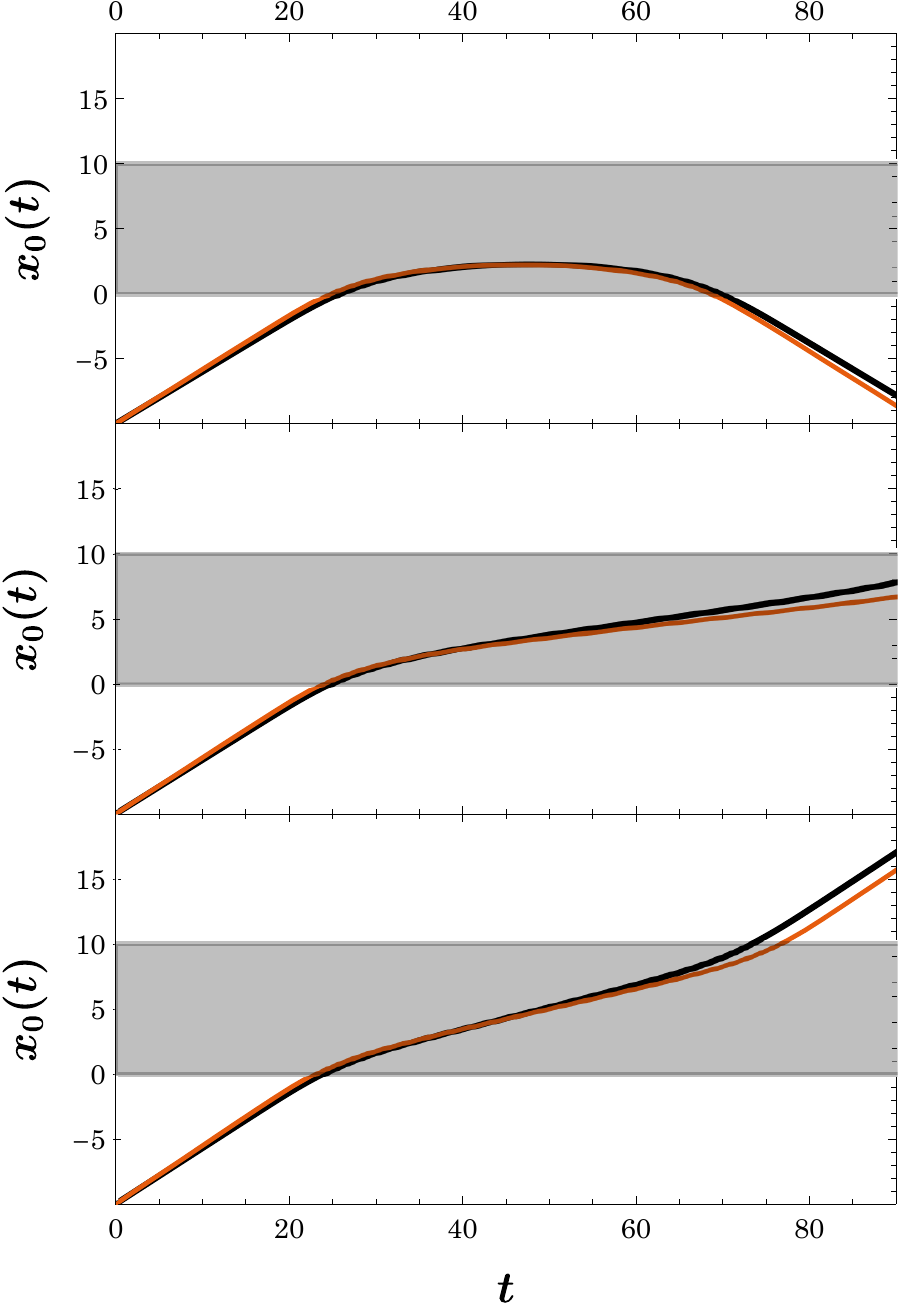}}}
    \quad
    \subfloat{{\includegraphics[width=7.5cm]{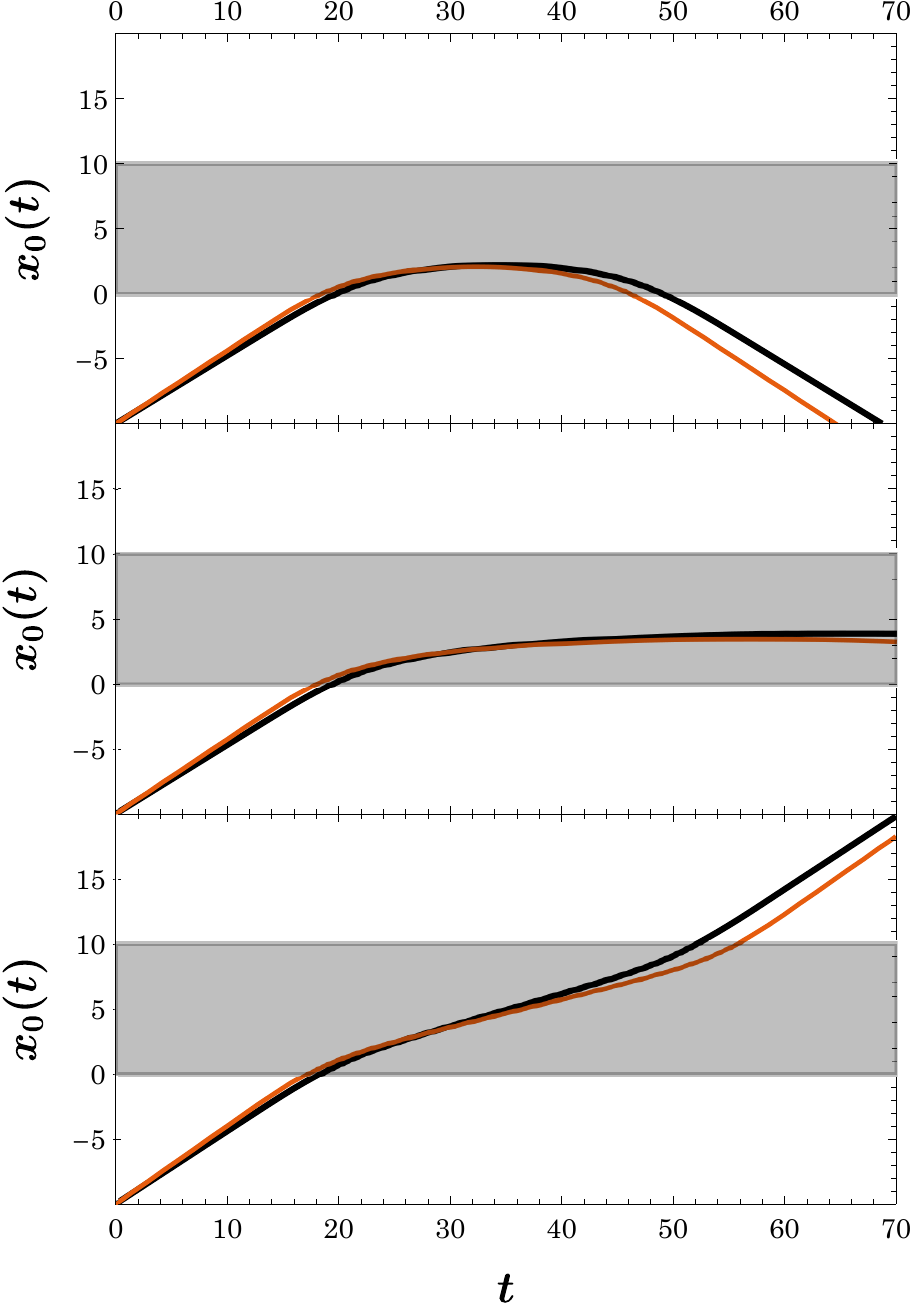}}}
    \caption{Comparison of the position of the center of mass of the kink for the solution of the original field  model (black line) and the model with two degrees of freedom (red line) for on the left $\varepsilon = 0.1$ and velocities from the top $0.4$, $0.415$, $0.43$, and on the right $\varepsilon = 0.2$ and velocities from the top $0.523$, $0.53$, $0.56$. }
    \label{fig_13}
\end{figure}

\begin{figure}[ht]
    \centering
    \includegraphics[width=9cm]{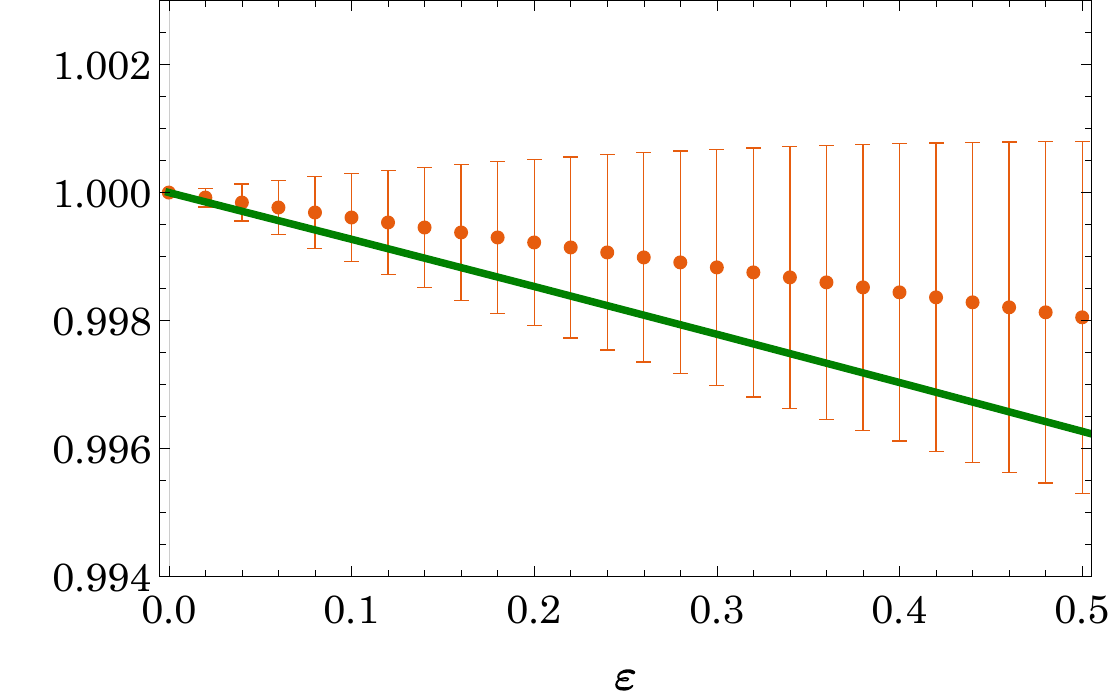}
    \caption{Fit of $\gamma$ values depending on the $\varepsilon$ determined from the solution of the original PDE (orange dots) with standard error calculated by maximum likelihood estimation compared with the $\gamma$ determined from a model with two degrees of freedom (green line). The figure describes the evolution when the inhomogeneity is in the form of a combination of hyperbolic tangents. In addition, $\alpha=0$ and $\Gamma=0$.}
    \label{fig_14}
\end{figure}

\begin{figure}[ht]
    \centering
    \subfloat{{\includegraphics[width=7.5cm]{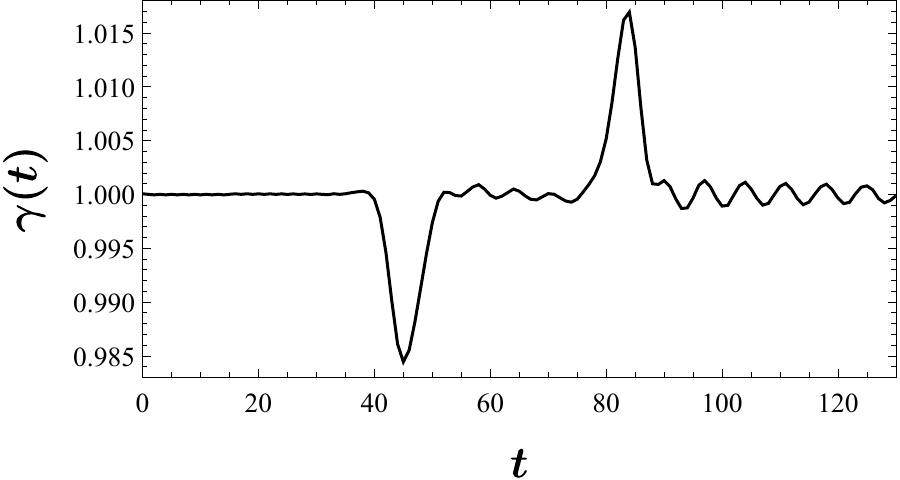}}}
    \quad
    \subfloat{{\includegraphics[width=7.5cm]{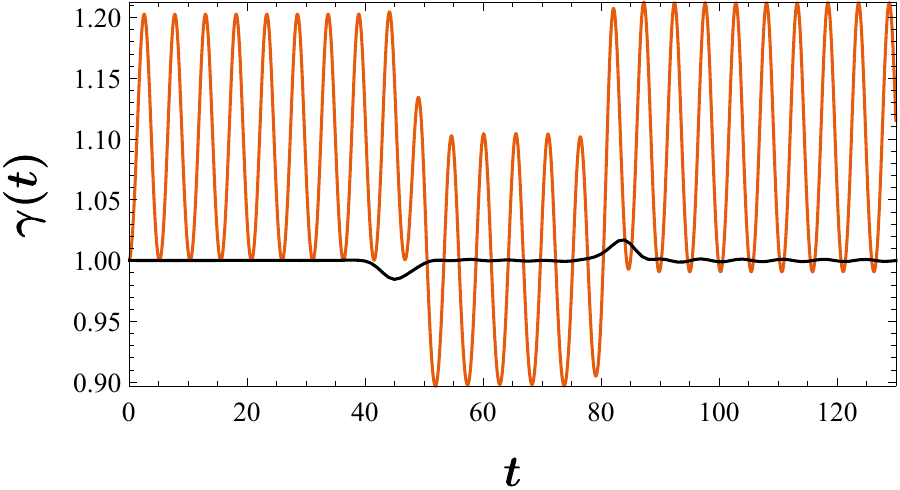}}}
    \caption{
    In the left panel of the figure, the evolution of the $\gamma(t)$ variable in the original field model. The right panel of the figure compares the results of the field model with an effective model with two degrees of freedom. In both cases $\varepsilon=0.1$.}
    \label{fig_15}
\end{figure}

The trajectory describing the movement of the center of mass resulting from the equation \eqref{sine-gordon} (black line) is compared with the time dependence of the collective variable $x_0(t)$ (red line) in figure \ref{fig_13}. In the figure, very good agreement between the effective model and the field model is achieved up to $\varepsilon$ values equal to $0.2$. In the figures, the parameter $L$ describing the width of the inhomogeneity is equal to $10$. It is interesting to note that the introduction of a second variable $\gamma(t)$ significantly improves the predictions of the effective model relative to the $x_0(t)$ variable. On the other hand, predictions about the $\gamma(t)$ variable itself are of more limited value. As a kink approaches the heterogeneity its width becomes suitably modulated. The changes in thickness gradually disappear with time at the field-theoretic level, after which the thickness stabilizes at a level characteristic of the stationary kink solution. Figure \ref{fig_14} compares the thickness of the static kink solution that follows from the effective model and the corresponding value derived from the field model. It can be seen that as $\varepsilon$ increases, the model increasingly underestimates the value of the $\gamma$ variable, although the relevant deviation is quite small and also it is clear that the model captures the nature of the qualitative trend of the effect of the perturbation of $\varepsilon$ on the parameter $\gamma$. Figure \ref{fig_15} compares the oscillations of the $\gamma$ variable in the effective model and the oscillations of the kink thickness as the kink passes through the inhomogeneity. The differences here are nontrivial although in both descriptions (i.e., exact and effective) the nature of the vibration changes similarly in the area of inhomogeneity. At the level of Eqs.~(\ref{2dof_oem}), one can trace this effect in the presence of terms such as the one $\propto \gamma^2 \dot{x}^2$ in the dynamical equation for the evolution of $\gamma(t)$. Indeed, while we observe that the field dynamics retain $\gamma$ to a nearly constant value far from the inhomogeneity, the above mentioned term is ``active'' in the reduced model equation leading to oscillatory dynamics of the kink width. Indeed, this is a point of potential future improvement of the reduced model as the latter is not presently capturing the Lorentz invariance of the homogeneous kink which would enable it to move with constant speed without inducing a width vibration. Nevertheless, the qualitative trends of variation of $\gamma(t)$ induced by the inhomogeneity are captured by the two-degree-of-freedom model (superimposed to the above mentioned vibration).

\begin{figure}[ht]
    \centering
    \includegraphics[width=9cm]{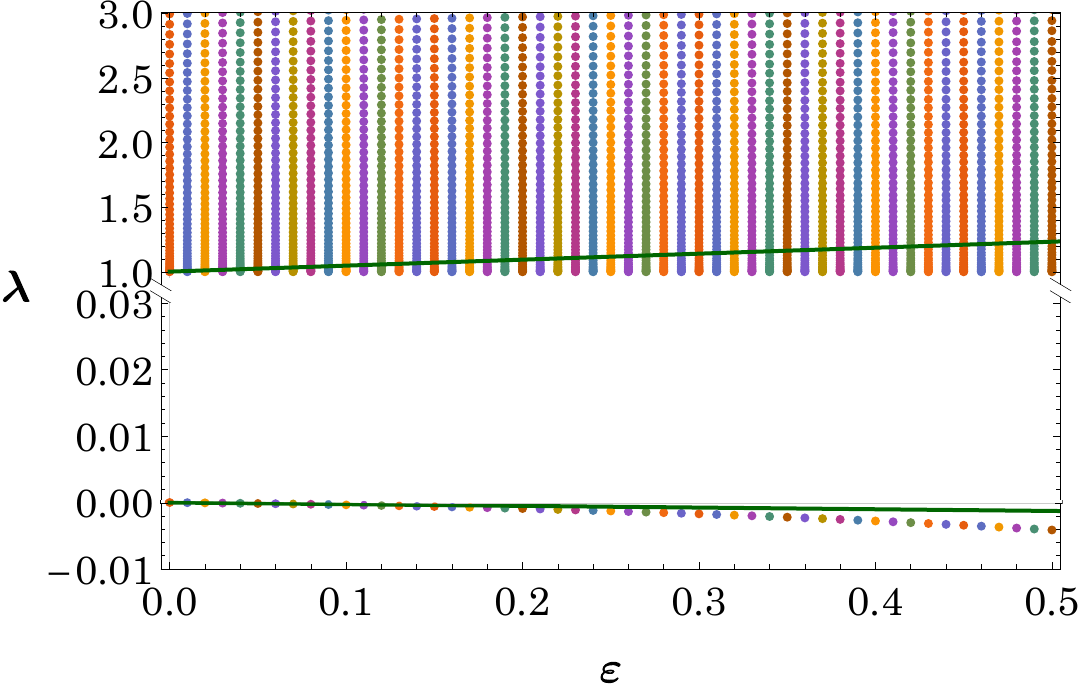}
    \caption{Comparison of the determined squared eigenfrequencies
    $\lambda=\omega^2$ from the
    linearization Jacobian (in case of $\alpha=0$, $\Gamma=0$) with those obtained from the model with two degrees of freedom linearized
    around the equilibrium state of the latter (green line).}
    \label{fig_16}
\end{figure}

Finally, the two-degree-of-freedom effective model naturally reproduces the two modes belonging to the spectrum of the $\hat{{\cal L}}$ operator (see figure \ref{fig_16}) in the case of the unstable saddle equilibrium of the Hamiltonian model kink centered at the impurity region. The first unstable mode (corresponding to the instability of the kink's position at the center of the inhomogeneity) is associated with the $x_0$ variable i.e., pertains to the former translational mode, and the oscillating mode (corresponding to changes in the thickness of the kink) is associated with the $\gamma$ variable. This second mode is essentially connected with the band edge of the continuous part of the spectrum of the operator $\hat{{\cal L}}$. Both modes of the effective model are represented by green lines. 

\subsubsection{The method of projecting onto the zero mode (for arbitrary $\alpha$ and $\Gamma$)}
In the case of two degrees of freedom, we insert $\xi (t,x)$ into the \eqref{model-2c} equation in the form of
$$
\xi (t,x) = \gamma(t) (x -x_0(t)) .
$$
This substitution results in the equation
\begin{equation}
\begin{gathered}
\label{zero-2a}
\left[ \left(\frac{\ddot{\gamma}}{\gamma} + \alpha
\frac{\dot{\gamma}}{\gamma}\right) \xi - \left( 2 \dot{x}_0
\dot{\gamma} + \alpha \gamma \dot{x}_0 + \gamma \ddot{x}_0 \right)
\right] \partial_{\xi} \phi_K + \left[ \left(
\frac{\dot{\gamma}}{\gamma} \right)^2 \xi^2 - 2 \dot{x}_0
\dot{\gamma} \xi  + \left(\gamma^2 \dot{x}_0^2 - \gamma^2 +1
\right) \right]
\partial^2_{\xi} \phi_K
\\
 =
\varepsilon \gamma (\partial_x \,g) \, \partial_{\xi} \phi_K +
\varepsilon \gamma^2 g \,   \partial_{\xi}^2 \phi_K  - \Gamma \, .
\end{gathered}
\end{equation}
The first of the equations of the two-degree-of-freedom effective model is obtained similarly to the one-degree-of-freedom model, i.e. by projecting to the zero mode. We multiply the above equation by the derivative of the kink ansatz, and then perform an integration 
over the entire domain to remove the dependence on the spatial variable.
In the case of the second equation, before calculating the integrals, we additionally multiply the equation by $\xi$, which changes the parity of the calculated integrals
\begin{equation}
\begin{gathered}
\label{zero-2b}
    \Ddot{x}_0+ \alpha \dot{x}_0+\frac{\Dot{\gamma}}{\gamma}\Dot{x}_0 -\frac{1}{8}\varepsilon\gamma \mathcal{J}=\frac{\pi}{4 \gamma} \Gamma, \\
    \frac{2 \pi^2}{3} \left(\frac{\Ddot{\gamma}}{\gamma} + \alpha \frac{\dot{\gamma}}{\gamma}\right) - \pi^2 \frac{\Dot{\gamma}^2}{\gamma^2}-4 \gamma^2 \Dot{x}^2+4\left(\gamma^2-1\right) - \varepsilon \gamma \mathcal{I}_1 - \varepsilon \gamma^2 \mathcal{I}_2=0.
\end{gathered}
\end{equation}
The integrals appearing in the equations for different forms of the function $g$ describing the inhomogeneity have the form
\begin{equation}\label{integral}
  \mathcal{J}=\int_{-\infty}^{+\infty} g(x) (\partial_{\xi} \phi_K )
  (\partial^2_{\xi} \phi_K ) d \xi , \,\,\,  \mathcal{I}_1=\int_{-\infty}^{+\infty} (\partial_x g(x) ) ~\xi ~(\partial_{\xi} \phi_K )^2
  d \xi
   , \,\,\,  \mathcal{I}_2=\int_{-\infty}^{+\infty} g(x)  ~\xi ~(\partial_{\xi} \phi_K ) (\partial^2_{\xi} \phi_K )
  d \xi .
\end{equation}
In the simplest case where the function $g$ consists of step functions i.e. $g(x) = \theta(x) -\theta(x-L)$, the equations of motion can be converted to the form
\begin{equation}
\begin{gathered}
\label{2dof_oem+}
    \Ddot{x}+ \alpha \dot{x}_0+\frac{\Dot{\gamma}}{\gamma}\Dot{x}_0+\frac{1}{4}\varepsilon\gamma \left[\sech^2(\gamma x_0) -\sech^2(\gamma(x_0-L)) \right]= \frac{\pi}{4 \gamma} \Gamma, \\
    \frac{2 \pi^2}{3} \left(\frac{\Ddot{\gamma}}{\gamma} + \alpha \frac{\dot{\gamma}}{\gamma} \right) - \pi^2 \frac{\Dot{\gamma}^2}{\gamma^2}-4 \gamma^2 \Dot{x}^2+4\left(\gamma^2-1\right) + 2\varepsilon \gamma^2\left[\tanh(\gamma x_0)-\tanh(\gamma (x_0-L))\right]\\+ 2\varepsilon \gamma^3 \left[x_0 \sech^2(\gamma x_0)-(x_0-L) \sech^2(\gamma(x_0-L))\right]=0.
\end{gathered}
\end{equation}
Note that these equations reduce to the  system \eqref{2dof_oem} when we take $\alpha$ and $\Gamma$ equal to zero.

\subsubsection{The method based on non-conservative lagrangian (for arbitrary $\alpha$ and $\Gamma$)}
\begin{figure}[ht]
    \centering
    \subfloat{{\includegraphics[width=7.5cm]{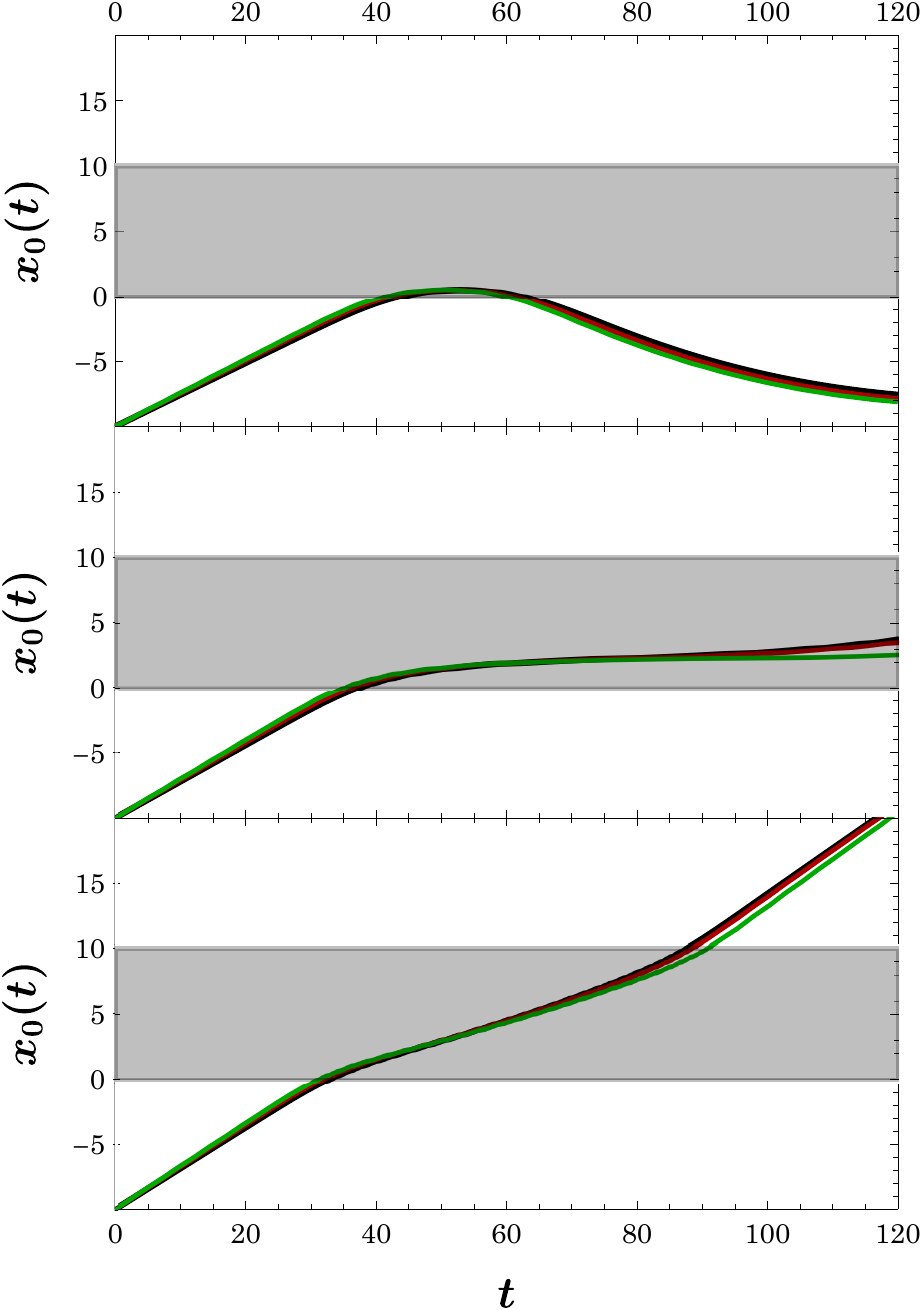}}}
    \quad
    \subfloat{{\includegraphics[width=7.35cm]{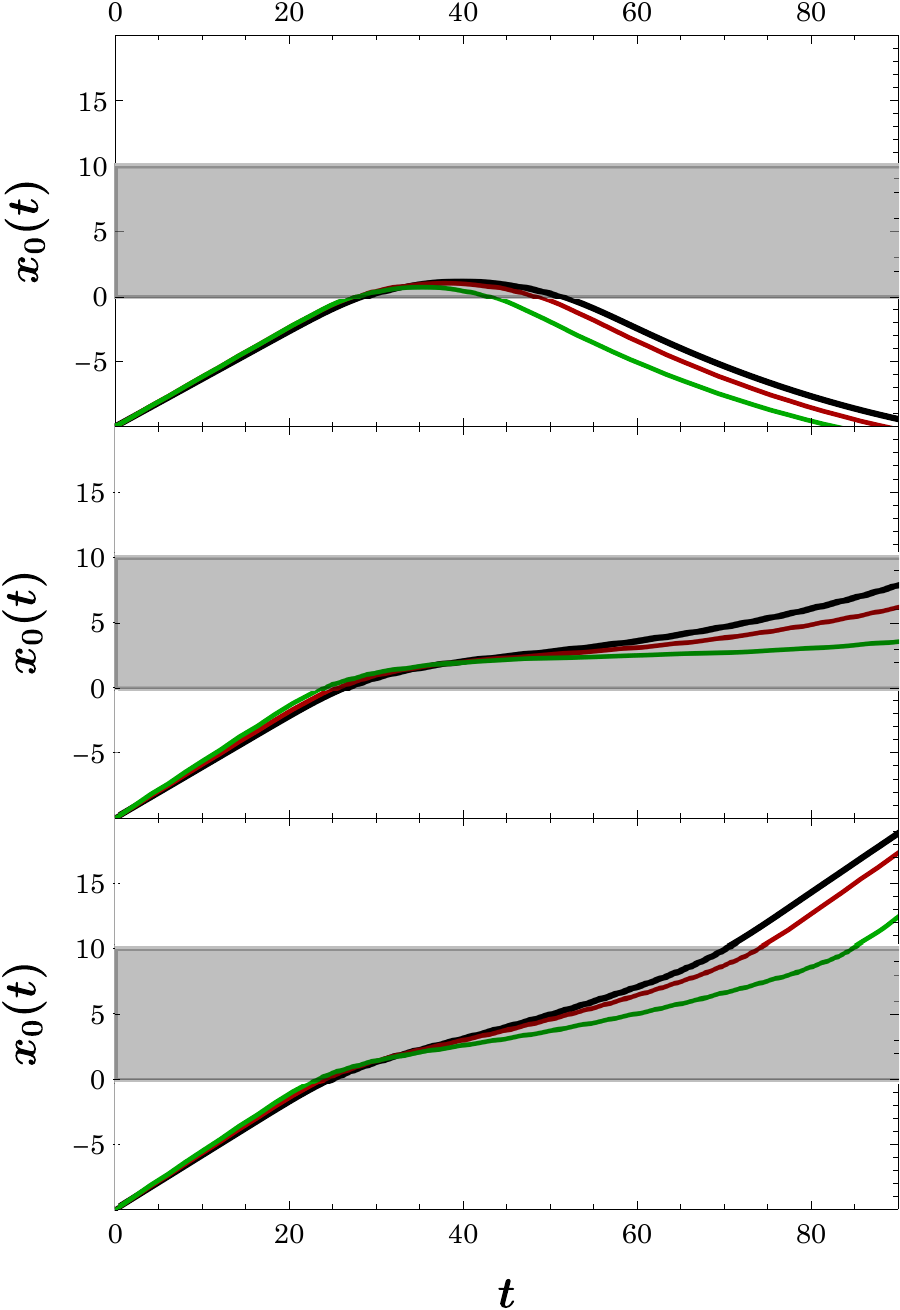}}}
    \caption{Comparison of the position of the center of mass of the kink for the solution from the original field model (black line) and models with two degrees of freedom based on projecting onto the zero mode according to \eqref{zero-2b} (green line) and on non-conservative Lagrangian according to \eqref{model_2st_3-c} (red line). The left panel contains figures for inhomogeneity strength $\varepsilon=0.05$ and bias current with values starting from the top $0.0032$, $0.0037$ and $0.0042$. On the right panel, results for $\varepsilon=0.1$ are presented. Starting from the top, the bias current is $0.005$, $0.0054$ and $0.0058$. In each case, the dissipation was equal to $0.01$.}
    \label{fig_17}
\end{figure}

As we described in the previous sections, Equation  \eqref{sine-gordon} can be obtained using a non-conservative Lagrangian density through equation  \eqref{model-3a}, where $\mathcal{R}$ represents the non-conservative contribution. In the case of an effective model with two degrees of freedom, the effective conservative Lagrangian and the effective non-conservative potential are obtained by integrating over the spatial variable \eqref{model-3b}. The only difference is the assumed ansatz, which in the case of two degrees of freedom has the form described by  equation  \eqref{an_2dof}.
In the model defined in this way, we have two effective equations of motion
\begin{equation}
\begin{gathered}
\label{model_2st_3-a} \partial_t \left( \frac{\partial L}{\partial
\dot{x}_0}\right)  - \frac{\partial L}{\partial x_0} =
\left[ \frac{\partial R}{\partial x_{-}} - \partial_t \left(
\frac{\partial R}{\partial \dot{x}_{-}}\right) \right]_{PL} , 
\\
\partial_t \left( \frac{\partial L}{\partial
\dot{\gamma}}\right)  - \frac{\partial L}{\partial \gamma} =
\left[ \frac{\partial R}{\partial \gamma_{-}} - \partial_t \left(
\frac{\partial R}{\partial \dot{\gamma}_{-}}\right) \right]_{PL} .
\end{gathered}
\end{equation}
In the physical limit (denoted here by $PL$) the auxiliary variables $x_{-}$ and $\gamma_{-}$ disappear, while $x_{+} \rightarrow x_0$ and $\gamma_{+} \rightarrow \gamma$. On the other hand, the Lagrangian consists of the sine-Gordon part \eqref{lan_sg_val_2dof} and 
the interaction term i.e. $L=L_{SG}+L_{\varepsilon}$ and therefore the equations of motion can be written as follows
\begin{equation}
\begin{gathered}
\label{model_2st_3-b} \Ddot{x}_0 + \frac{\dot{\gamma}}{\gamma} \dot{x}_0 - \frac{1}{8 \gamma} \frac{\partial L_{\varepsilon}}{\partial x_0} = \frac{1}{8 \gamma}
\left[ \frac{\partial R}{\partial x_{-}} - \partial_t \left(
\frac{\partial R}{\partial \dot{x}_{-}}\right) \right]_{PL} , 
\\
\frac{2 \pi^2}{3} \frac{\Ddot{\gamma}}{\gamma} - \pi^2 \frac{\dot{\gamma}^2}{\gamma^2} - 4 \gamma^2 \dot{x}^2_0 + 4 (\gamma^2-1) - \gamma^2 \frac{\partial L_{\varepsilon}}{\partial \gamma} =
\gamma^2 \left[ \frac{\partial R}{\partial \gamma_{-}} - \partial_t \left(
\frac{\partial R}{\partial \dot{\gamma}_{-}}\right) \right]_{PL} .
\end{gathered}
\end{equation}
The right-hand sides of these equations we obtain by calculating the non-conservative potential and its derivatives in the physical limit
\begin{equation}
\begin{gathered}
\label{model_2st_3-c} \Ddot{x}_0 + \frac{\dot{\gamma}}{\gamma} \dot{x}_0 - \frac{1}{8 \gamma} \frac{\partial L_{\varepsilon}}{\partial x_0} = - \alpha \dot{x}_0 + \frac{\pi}{4 \gamma} \Gamma , 
\\
\frac{2 \pi^2}{3} \frac{\Ddot{\gamma}}{\gamma} - \pi^2 \frac{\dot{\gamma}^2}{\gamma^2} - 4 \gamma^2 \dot{x}^2_0 + 4 (\gamma^2-1) - \gamma^2 \frac{\partial L_{\varepsilon}}{\partial \gamma} =
- \frac{2 \pi^2}{3} \alpha \frac{\dot{\gamma}}{\gamma} .
\end{gathered}
\end{equation}
We recall here that the interaction with the inhomogeneity is described by the following integral
\begin{equation}
    L_{\varepsilon} = - \frac{1}{2} \varepsilon \int_{-\infty}^{+\infty} g(x) (\partial_x \phi_K)^2,
\end{equation}
where $\phi_K$ denotes the ansatz  \eqref{an_2dof}.
For example, in the case of inhomogeneity defined by the unit step functions 
$g(x) = \theta(x) - \theta(x-L)$ we get equations identical to the formulas \eqref{2dof_oem+}, obtained previously using the
projection approach. 
A different situation occurs for inhomogeneities defined by
hyperbolic tangents $g(x)=\tanh{(x)} -\tanh{(x-L)}$.
Naturally, if in the system there is  dissipation, we can only use the method based on the non-conservative Lagrangian and the method of projection onto the zero mode. The results of the two methods are found
to differ slightly as shown in Figure \ref{fig_17}, favoring the non-conservative Lagrangian method as more accurate. Figure \ref{fig_17} compares trajectories obtained for the inhomogeneity of $g(x)=\tanh(x) - \tanh(x-L)$ using a method based on a non-conservative Lagrangian for two degrees of freedom (red line) and a zero mode projection (green line). In all figures, the black line represents the result obtained from the
full field model PDE. The inhomogeneity in the figures of the left panel corresponds to $\varepsilon=0.05$, while for the right panel it is $\varepsilon=0.1$. We used the following currents on the left panel, starting from the top, 0.0032, 0.0037 and 0.0042. On the right panel, the currents are assumed to be (again, starting from the top) 0.005, 0.0054 and 0.0058. In all figures, the dispersion is assumed to be $\alpha=0.01$. We found that as the parameter $\varepsilon$ increases, the method of projecting onto the zero mode to a higher extent than the method based on the non-conservative Lagrangian underestimates the position of the kink.

\section{Conclusions and Future Work}
In this paper, we have studied the behavior of the kink in the sine-Gordon model in the presence of a localized inhomogeneity. In the case without dissipation, we focused on the interaction of the kink with the impurity region at speeds proximal to the critical velocity separating transmission from reflection. In the immediate vicinity of the relevant critical point (which we identified as a saddle), we observed the kink slowing down for an extended time interval at the center of the inhomogeneity. The process of the kink interaction with the inhomogeneity was also described within the framework of effective models with one and two degrees of freedom. As expected, the description with one collective variable works well for small values of the perturbation parameter $\varepsilon$. On the other hand, the inclusion of a second collective variable, effectively characterizing the width of the coherent structure, significantly improves the predictions of the effective model including for somewhat larger values of $\varepsilon$. At the same time, however, the predictions for the second collective variable bear some diferences in comparison to the field description albeit in ways that were explained in the associated discussion. In particular, the second collective variable is intended to identify the occurrence of the interaction, while the reduced description also seems to identify a vibrational mode associated with the edge of the continuous spectrum. A more refined representation of the relevant mode that yields a close agreement with the field-theoretic results constitutes a natural question for future study.

The case of interaction of a kink with inhomogeneity in a system with dissipation and external drive presents further intriguing features in its own right. The passage of the kink through the barrier or reflection depends on the relationship between the external forcing and the dissipation. Particularly interesting here is the process of interaction of the kink with the barrier for bias currents smaller than the threshold current. In this case, we observe successive reflections of the kink from the barrier caused by the bias current pushing it toward the barrier. Oscillations of the kink position are naturally damped due to the presence of dissipation in the system. The shape of the final static configuration can be determined on the basis of a linearized approximation and reveals a stable spiral fixed point of the effective description. The reduced model description of the process of interaction of the kink with the inhomogeneity below the threshold current leads to surprisingly consistent results with those from the original PDE. Moreover, the effective models with two degrees of freedom (for small $\varepsilon$) correctly approximate the excitation spectrum obtained on the ground of the linear approximation (fig. \ref{fig_16}), as well as the associated dynamics. While both related methods are found to be qualitatively adequate, the non-conservative Lagrangian approach developed herein is also found to be highly quantitatively accurate in describing the kink evolution. [For one degree of freedom, the different approaches developed are found to yield identical results in suitable limits].

This study naturally paves the way for a number of future possibilities. On the one hand, our focus here was in the interaction of sine-Gordon kink with an inhomogeneity. Yet, there has been a rich literature exploring the resonant interaction of a $\phi^4$ kink with an impurity dating back to the work of~\cite{feizhang} that has recently seen a resurgence of interested in different model variants and associated phenomenologies~\cite{spectralwall,Lizunova2021}. Another interesting direction concerns the exploration of higher dimensional variants even of the sine-Gordon variety, in order to appreciate the effects of curvature and impurity geometry on the kink dynamics; see for some recent examples~\cite{Kevrekidis2018,Gonzalez2022}. Such studies are currently in progress and will be reported in future publications.

\section{Appendix: Equivalence of the first two approaches for
one degree of freedom reduced models} In this section, we will consider a class of Lagrangian systems, in the absence of dissipation for which we will demonstrate the equivalence of the first two approaches presented in the main body of the text for one degree of freedom effective descriptions for the wave's center of mass. We assume that stable solutions in the form of solitons are present in this system. The effective description of the position of the soliton in this system is specified by the equation:
\begin{equation}\label{ap1}
\frac{\partial L_{eff}}{\partial x_0} - \frac{d}{dt} \left(
\frac{\partial L_{eff}}{\partial \dot{x}_0} \right) = 0.
\end{equation}
Here $L_{eff}$ denotes the effective lagrangian
\begin{equation}\label{ap2}
L_{eff} = \int_{-\infty}^{+\infty}
\mathcal{L}(\phi_K,\dot{\phi}_K,\phi_K') dx . 
\end{equation}
As can be seen, it is obtained by integrating the Lagrangian density of the underlying field theory with respect to the spatial variable. Here $\phi_K = \phi_K(x-x_0(t))$ denotes the kink solution and $x_0(t)$ is the collective variable that represents the position of the kink. We will consider Lagrangians that contain terms that explicitly break the translational symmetry of the system. An example Lagrangian of this type has the form:
\begin{equation}\label{ap3} \mathcal{L}(\phi_K,\dot{\phi}_K,\phi_K') = \frac{1}{2} \, \mathcal{A}(x) \,
\dot{\phi}_K^2 - \frac{1}{2} \, \mathcal{F}(x) \, {\phi_K'}^2 -
\mathcal{B}(x) V(\phi_K) ,
\end{equation}
where the dot denotes the time derivative and prime denotes the derivative with respect to space variable $x$. Here $\mathcal{A}$, $\mathcal{B}$ and $\mathcal{F}$ are arbitrary functions with finite values. In particular, in previous sections of this work, we considered the cases for which $\mathcal{A}=1$ and $\mathcal{B}=1$. Applying the definition of the effective Lagrangian \eqref{ap2} to the equation \eqref{ap1} we obtain: 
\begin{equation}\label{ap4} 
\int_{-\infty}^{+\infty} \left\{ \frac{\partial
\mathcal{L}}{\partial x_0} - \frac{d}{dt} \left( \frac{\partial
\mathcal{L}}{\partial \dot{x}_0} \right)\right\} dx =0
\end{equation}
When calculating the derivatives, we must remember that the Lagrangian density depends on the $x_0$ variable both through the field $\phi_K$, its spatial and time derivatives
\begin{equation}\label{ap5} 
\frac{\partial \mathcal{L}}{\partial x_0} =  \frac{\partial
\mathcal{L}}{\partial \phi_K} \, \frac{\partial \phi_K}{\partial
x_0}+ \frac{\partial \mathcal{L}}{\partial \dot{\phi}_K} \,
\frac{\partial \dot{\phi}_K}{\partial x_0} + \frac{\partial
\mathcal{L}}{\partial \phi'_K} \, \frac{\partial \phi'_K}{\partial
x_0} .
\end{equation}
Next, we will convert the derivatives with respect to the variable describing the position of the kink $x_0$  into derivatives with respect to the variable $\xi=x-x_0(t)$
\begin{equation}\label{ap6} 
\frac{\partial \mathcal{L}}{\partial x_0} =  - \frac{\partial
\mathcal{L}}{\partial \phi_K} \, \frac{\partial \phi_K}{\partial
\xi}- \frac{\partial \mathcal{L}}{\partial \dot{\phi}_K} \,
\frac{\partial \dot{\phi}_K}{\partial \xi} - \frac{\partial
\mathcal{L}}{\partial \phi'_K} \, \frac{\partial \phi'_K}{\partial
\xi} .
\end{equation}
Similarly, we can calculate the derivative with respect to the kink velocity, but this time the dependence on the $\dot{x}_0$ variable occurs in just one term
\begin{equation}\label{ap7}
\frac{\partial \mathcal{L}}{\partial \dot{x}_0} = \frac{\partial
\mathcal{L}}{\partial \phi_K} \, \frac{\partial \phi_K}{\partial
\dot{x}_0}+ \frac{\partial \mathcal{L}}{\partial \dot{\phi}_K} \,
\frac{\partial \dot{\phi}_K}{\partial \dot{x}_0} + \frac{\partial
\mathcal{L}}{\partial \phi'_K} \, \frac{\partial \phi'_K}{\partial
\dot{x}_0} =\frac{\partial \mathcal{L}}{\partial \dot{\phi}_K} \,
\frac{\partial \dot{\phi}_K}{\partial \dot{x}_0} .
\end{equation}
The derivative of the field $\phi_K$ with respect to time explicitly depends in a linear way on the velocity
\begin{equation}\label{ap8}
\dot{\phi}_K = \frac{\partial \phi_K}{\partial \xi} \, \frac{d
\xi}{d t} = - \dot{x}_0 \frac{\partial \phi_K}{\partial \xi}
\end{equation}
and therefore the derivative of $\dot{\phi}_K$ with respect to velocity is equal to
\begin{equation}\label{ap9}
\frac{\partial \dot{\phi}_K}{\partial \dot{x}_0} = -
\frac{\partial \phi_K}{\partial \xi} .
\end{equation}
The relevant contribution of the derivative of the Lagrangian with respect to $\dot{x}_0$ then yields:
\begin{equation}\label{ap10}
\frac{\partial \mathcal{L}}{\partial \dot{x}_0} = - \frac{\partial
\mathcal{L}}{\partial \dot{\phi}_K} \, \frac{\partial
\dot{\phi}_K}{\partial \xi}.
\end{equation}
The obtained derivatives of Lagrangian density with respect to $x_0$, given by the equation \eqref{ap6} and $\dot{x}_0$, set by the equation \eqref{ap10} can be used in equation \eqref{ap4} yielding
\begin{equation}\label{ap11}
\int_{-\infty}^{+\infty} \left\{ - \frac{\partial
\mathcal{L}}{\partial \phi_K} \, \frac{\partial \phi_K}{\partial
\xi}- \frac{\partial \mathcal{L}}{\partial \dot{\phi}_K} \,
\frac{\partial \dot{\phi}_K}{\partial \xi} - \frac{\partial
\mathcal{L}}{\partial \phi'_K} \, \frac{\partial \phi'_K}{\partial
\xi}
 + \frac{d}{dt} \left(
\frac{\partial \mathcal{L}}{\partial \dot{\phi}_K} \,
\frac{\partial \phi_K}{\partial \xi }\right) \right\} dx =0 .
\end{equation}
After performing the differentiation with respect to time, we can separate the part that is multiplied by the zero mode and the other part that we still need to transform
\begin{equation}\label{ap12}
\int_{-\infty}^{+\infty} \left\{ \frac{d}{dt} \left(
\frac{\partial \mathcal{L}}{\partial \dot{\phi}_K} \right) -
\frac{\partial \mathcal{L}}{\partial \phi_K} \right\} \frac{\partial
\phi_K}{\partial \xi } d \xi - \int_{-\infty}^{+\infty}
\left\{ \frac{\partial \mathcal{L}}{\partial \phi'_K} \,
\frac{\partial \phi'_K}{\partial \xi}
   \right\} d x =0 .
\end{equation}
In the second integral, we transfer the derivative with respect to the spatial variable from the the second factor to the first one
\begin{equation}\label{ap13}
\int_{-\infty}^{+\infty}  \left\{ \frac{\partial
\mathcal{L}}{\partial \phi'_K} \, \frac{\partial \phi'_K}{\partial
\xi}
   \right\} d x =\int_{-\infty}^{+\infty}  \frac{\partial}{\partial x}\left\{ \frac{\partial
\mathcal{L}}{\partial \phi'_K} \,  \left( \frac{\partial
\phi_K}{\partial \xi} \right)
   \right\} d x - \int_{-\infty}^{+\infty}  \frac{\partial}{\partial x}\left( \frac{\partial
\mathcal{L}}{\partial \phi'_K} \right) \,   \frac{\partial
\phi_K}{\partial \xi}
   d x .
\end{equation}
Since the spatial derivative of $\phi_K$ vanishes at infinity so we can extract the term that contains multiplication by the zero mode
\begin{equation}\label{ap14}
\int_{-\infty}^{+\infty}  \left\{ \frac{\partial
\mathcal{L}}{\partial \phi'_K} \, \frac{\partial \phi'_K}{\partial
\xi}
   \right\} d x = - \int_{-\infty}^{+\infty}  \frac{\partial}{\partial x}\left( \frac{\partial
\mathcal{L}}{\partial \phi'_K} \right) \,   \frac{\partial
\phi_K}{\partial \xi} \,
   d \xi .
\end{equation}
The integral that is transformed in this way can be reinserted into equation \eqref{ap12} yielding
\begin{equation}\label{ap15}
\int_{-\infty}^{+\infty} \left\{ \frac{d}{dt} \left(
\frac{\partial \mathcal{L}}{\partial \dot{\phi}_K} \right) +
\frac{\partial}{\partial x}\left( \frac{\partial
\mathcal{L}}{\partial \phi'_K} \right) - \frac{\partial
\mathcal{L}}{\partial \phi_K} \right\} \frac{\partial
\phi_K}{\partial \xi } \, d \xi  =0 .
\end{equation}
In relativistic notation, the last equation can be written as follows
\begin{equation}\label{ap16}
\int_{-\infty}^{+\infty} \left\{ \partial_{\mu} \left(
\frac{\partial \mathcal{L}}{\partial ( \partial_{\mu} \phi_K )}
\right)  - \frac{\partial \mathcal{L}}{\partial \phi_K} \right\}
\frac{\partial \phi_K}{\partial \xi } \, d \xi  = 0 ,
\end{equation}
where $\partial_{\mu}$ denotes differentiation with respect to space-time variables $x^{\mu}=(x^0,x^1)=(t,x)$. Note that starting from equation \eqref{ap1} we obtained equation \eqref{ap16}, which defines the method of projecting onto the zero mode. On the other hand, going backwards in our calculations from the equation \eqref{ap16}, we arrive at the effective equation \eqref{ap1}, which means that the method based on the conservative Lagrangian is equivalent (in the absence of dissipation) to the method of zero mode projection. Obviously, our considerations apply to the effective model with one degree of freedom, yet the relevant proof applies for arbitrary nonlinearity described by $V(\phi)$ and arbitrary form of the heterogeneity in the model.

Finally, let us also notice that naturally, in the case where $\alpha=0$ and $\Gamma=0$, generally (at the level of field equations as well as effective equations) the approach based on the non-conservative Lagrangian is equivalent to the approach based on the conservative Lagrangian. In this case the non-conservative contribution $R$ is equal to zero, so the equation \eqref{model-3c} reduces to the equation \eqref{ap1}, showing the equivalence of the two models. Accordingly, in this case, all three approaches are equivalent.

\section*{Acknowledgement}
This research has been made possible by the Kosciuszko Foundation. The American Centre of Polish Culture. 
This material is based upon work supported by the U.S.\ National Science Foundation under the awards PHY-2110030 and DMS-2204702 (PGK).

\printbibliography

\end{document}